\documentclass[a4paper,fleqn,usenatbib]{mnras}

\usepackage{newtxtext,newtxmath}

\usepackage[T1]{fontenc}
\usepackage{ae,aecompl}

\usepackage{graphicx}	
\usepackage{amsmath}	
\usepackage{amssymb}	
\usepackage{natbib}

\newcommand{\beq}{\begin{equation}}
\newcommand{\eeq}{\end{equation}}

\def\asec{\ifmmode ^{\prime\prime}\else$^{\prime\prime}$\fi}

\def\msun{\hbox{M$_{\odot}$}}

\def\msunyr{\mbox{\,${\rm M_{\odot}\, yr^{-1}}$}}
\def\mdot{\dot M}

\def\degs{\ifmmode ^{\circ}\else$^{\circ}$\fi}
\def\amin{\ifmmode ^{\prime}\else$^{\prime}$\fi}
\def\asec{\ifmmode ^{\prime\prime}\else$^{\prime\prime}$\fi}

\def\degs{\ifmmode ^{\circ}\else$^{\circ}$\fi}
\def\amin{\ifmmode ^{\prime}\else$^{\prime}$\fi}

\def\EE#1{\times 10^{#1}}

\def\cm{\mbox{\,cm}}
\def\ccc{\mbox{\,cm$^{-3}$}}
\def\cm3{\mbox{\,cm$^{-3}$}}
\def\kms{\mbox{\,km~s$^{-1}$}}

\def\kms{\mbox{\,km s$^{-1}$}}

\def\lsim{\!\!\!\phantom{\le}\smash{\buildrel{}\over
 {\lower2.5dd\hbox{$\buildrel{\lower2dd\hbox{$\displaystyle<$}}\over
                                 \sim$}}}\,\,}
\def\gsim{\!\!\!\phantom{\ge}\smash{\buildrel{}\over
{\lower2.5dd\hbox{$\buildrel{\lower2dd\hbox{$\displaystyle>$}}\over
                               \sim$}}}\,\,}

\def\araa{ARA\&A}             
\def\apj{ApJ}                 
\def\apjl{ApJ}                
\def\apjs{ApJS}               
\def\aap{A\&A}                
\def\mnras{MNRAS}             
\def\nat{Nature}              
\def\pasa{PASA}	

\title[The Impact of the Environment of White Dwarf Mergers on Fast Radio Bursts]{The Impact of the Environment of White Dwarf Mergers on Fast Radio Bursts}

\author[]{
Esha Kundu$^{1}$\thanks{E-mail: esha.kundu@curtin.edu.au}, 
Lilia Ferrario$^{1,2}$
\\
$^{1}$International Centre for Radio Astronomy Research, Curtin University, Bentley, WA 6102, Australia\\
$^{2}$Mathematical Sciences Institute, The Australian National University, Canberra, ACT 2601, Australia
}
\date{Accepted XXX. Received YYY; in original form ZZZ}

\pubyear{2019}

\begin{document}
\label{firstpage}
\pagerange{\pageref{firstpage}--\pageref{lastpage}}
\maketitle

\begin{abstract}
Fast radio bursts (FRBs) are transient intense radio pulses with duration of milliseconds. Although the first FRB was detected more than a decade ago, the progenitors of these energetic events are not yet known. The currently preferred formation channel involves the formation of a neutron star~(NS)/magnetar. While these objects are often the end product of the core-collapse (CC) explosion of massive stars, they could also be the outcome of the merging of two massive white dwarfs. 
In the merger scenario the ejected material interacts with a constant-density circumbinary medium and creates supersonic shocks. We found that when a radio pulse passes through these shocks the dispersion measure (DM) increases with time during the free expansion phase. The rotation measure (RM) displays a similar trend if the power-law index, $n$, of the outer part of the ejecta is $>6$. For $n = 6$ the RM remains constant during this phase.
Later, when the ejecta move into the Sedov-Taylor phase while the DM still increases, however, with a different rate, the RM reduces. This behaviour is somewhat similar to that of FRB~121102 for which a marginal increase of DM and a 10\% decrease of RM have been observed over time. These features are in contrast to the CC scenario, where the DM and RM contributions to the radio signal always diminish with time. 
\end{abstract}



\section{Introduction}
\label{sec:intro}

Fast radio bursts (FRBs) are bright radio pulses of unknown origins \citep{lorimer07, thornton_13}. The high dispersion measure (DM) associated with FRBs points toward a cosmological origin of these events. This is in fact true for FRB\,121102, FRB\,180924, FRB\,190523, FRB\,181112 which are localised at a redshift of 0.19 \citep{tendulkar17}, 0.32 \citep{bannister19}, 0.66 \citep{ravi19} and 0.48 \citep{prochaska19}, respectively. On the other hand, the large DM observed in some FRBs could be caused by these sources being embedded in a highly ionised medium. After emission the radiation passes through a number of different media which contribute to the total DM and RM. The contributors are: i.) the immediate surrounding of the FRB source, ii.) the host galaxy of the progenitor, iii.) the extragalactic medium through which the radiation propagates, and iv.) our Milky Way galaxy.  
The medium surrounding a FRB provides important information about the progenitor system. As \citet{piro16} and \citet{piro18} have pointed out, even if the immediate environment of repeating FRBs may not dominate their DM and RM, it will dominate their changes over time and thus such changes should be measurable over the years. Hence the detailed studies of the DM and RM of FRBs are essential to unravel the mysteries that still surround their origin.

\par
Among the various possible progenitors a neutron star (NS)/magnetar is one of the potential systems that can produce FRBs
\citep{popov13,lyubarsky14,katz16,beloborodov17,lyutikov19,metzger19,wadiasingh19}.
Although a NS/magnetar is  the expected end product of the core collapse explosion of massive stars  \citep[$>$ 8 \msun,][]{smartt09}, the merger of two massive WDs may also lead to the same outcome.  
However, double WD mergers are also the currently favoured channel to Type Ia supernova (SN Ia) \citep{hill00,maoz14,hill13}. Indeed, the numerical simulations of \citet{pakmor10,pakmor12} have shown that if the masses of the two WDs are $\geq 0.9~\msun$ with a mass ratio $\sim$ 1, their coalescence will likely produce a SN Ia. On the other hand, different configurations of merging WDs may not lead to an explosion but to a collapse and the formation of a NS/magnetar 
\citep[e.g.][]{saio85,shen12}.
Another proposal is that such mergers could produce an ultra-massive and fast rotating WD which could also be a potential source of FRBs \citep{kashiyama13}. During the formation of this compact object some of the material is ejected into the surrounding medium. The interaction of this expanding matter with the surrounding medium generates two shock waves; a forward shock moving outward into the circumstellar medium (CSM), and a reverse shock that recedes backward in mass coordinate. This is exactly what happens when supernova (SN) ejecta interact with their ambient medium. Since these shocks have temperatures more than a million Kelvin  \citep{chevalier82,vink11,kundu19}, which implies the presence of ionised particles in the shocked region, and as these are ideal places for magnetic field amplification to take place \citep{bykov13,caprioli14}, the shocked region contributes to the total DM and RM of the radio wave as it passes through it.

\par 
 The propagation of FRBs through CC remnants has already been studied extensively \citep[e.g.,][]{piro18,piro16,zhang17} under the simplifying assumption of constant density ejecta. Analytical solutions and numerical simulations have revealed that whilst the inner regions of SN ejecta are flat, beyond a certain radius they acquire a steep power law profile \citep{matzner99,pakmor12}. In this work we have considered realistic ejecta profiles for both CC and merger scenarios and found that in case of double WD mergers the evolution of the DM and RM associated with the radio waves are unique in nature and different from the CC case. Therefore, if a source of FRBs is embedded in debris generated by a CC explosion or double WD merger, the studies of the time evolution of DM and RM should help us identify their progenitors.

\section{Evolution of the Shocked shell}
\label{sec:ShoSheEvo}
Although the ejected material in double WD mergers is much smaller than in SNe (see section \ref{sebsec:Merger_WDs}), we assume that the density profile $\rho_{\rm ej}(r)$ of the ejecta is the same as that observed in SNe Ia and CC SNe, that is, it is constant for $r < r_{\rm brk}$,  where $r_{\rm brk}$ is the radius corresponding to the break velocity $v_{\rm brk}$, and is $\propto r^{-n}$ for $r > r_{\rm brk}$, where $n$ is the power-law index and $\sim$ 10 \citep{matzner99,kundu17}.

Depending on the type of progenitors, the ambient media around these systems are different. In the case of the CC of a massive star, previous mass loss episodes due to strong stellar winds from the progenitor are important for the shaping of the CSM \citep{mokiem07,vanloon05,nugis00}. 
The density of the surrounding medium, $\rho_{\rm csm}$, is expected to have the form $\rho_{\rm csm}= \mdot/4\pi v_w r^{2}$, when $\mdot/v_w$ is assumed to be a constant. Here $\mdot$ is the mass-loss rate from the pre-SN star and $v_w$ its expulsion speed.

\par 
Regardless of whether the merging of two massive WDs produces a thermonuclear runway or a collapse into a NS/magnetar, the CSM density around the WDs prior to the ensuing explosion or collapse must be the same. Radio and X-ray observations of the media around SNe Ia have found them to be tenuous with a particle density $\lsim 100 ~\ccc$ \citep{chomiuk16, margutti12, kundu17}, and favoured a density profile $\rho_{\rm csm}$ that is constant with the radius. Thus we assume $\rho_{\rm csm} = \mu m_p n_{\rm {ISM}}=A$ (a constant), where $\mu$ and $n_{\rm {ISM}}$ represent the mean atomic weight of the ambient medium and the particle density, respectively, and $m_p$ is the mass of a proton.
\par


The interaction of the power-law ejecta with an ambient medium characterised by $\rho_{\rm csm} = A r^{-s}$, where $s =$ 0 (2) and $A = \mu m_p n_{\rm {ISM}}$  ($\mdot/4\pi v_w)$, can be described by the self-similar solutions \citep{chevalier82}. 
Therefore, the time evolution of the contact discontinuity ($r_c$) for $n>5$ in the free expansion phase can be given as 
\beq
r_c^{{\rm FE}} = D^{\frac{1}{n-s}} t^{\frac{n-3}{n-s}}
\label{eq:rc_FE}
\eeq 
\citep{chevalier82}, where $D = \xi v_{\rm brk}^n/A$ and $\xi$ is a constant. The forward and reverse shock radii scale as $r_s = \alpha r_c$ and $r_{rev}= \beta r_c$ with $\alpha > 1$ and $\beta < 1$. 
For different values of $n$ and $s$ the $\alpha$, $\beta$ and $\xi$ can be found in \cite{chevalier82}.
The duration of the free expansion phase can be estimated as 
\beq
T_{\rm{FE}} = \Lambda^\frac{1}{(3-s)} ~ \zeta_2^\frac{n-s}{3-s} ~ v_{\rm brk}^\frac{s}{3-s},
\label{eq:freeexp}
\eeq
where 
\beq
\Lambda = \frac{4 \pi}{\theta}\left(\frac{3-s}{n-3} \right)\frac{\rho_{ej,in}}{\Delta} \frac{\zeta_2^{3-n}}{\zeta_1^{3-s}} ~ r_{\rm o,w}^{2-s}
\eeq
\citep{kundu17}. Here $\Delta = \mdot/v_w$ ($n_{\rm ISM}$) for a wind (constant density) medium. The quantity $\theta$ is the ratio between the swept up ejecta and the CS mass, $r_{\rm o,w}$ is a reference radius, $\rho_{\rm ej,in}$ is the density of the inner (flat) ejecta and $\zeta_1 = r_s/r_c$ and $\zeta_2 = r_{rev}/r_c$. The free expansion ends when the reverse shock starts to penetrate into the inner part of the flat ejecta. The shocked shell then evolves into the Sedov-Taylor (ST) phase.  In this phase the shocks radii are $\propto t^{2/(5-s)}$ \citep{ostriker88,draine11}. The evolution of the contact discontinuity in the ST phase is
\beq
r_c^{\rm ST} = D^{\frac{1}{n-s}} ~ T_{\rm FE}^{\frac{(n-5)(3-s)}{(n-s)(5-s)}} ~ t^{\frac{2}{5-s}},
\label{eq:rc_ST}
\eeq 
where we assume that $r_c$ is continuous across $T_{\rm FE}$. It is presumed in this work that in the ST phase the same $\alpha$ and $\beta$ are applicable in calculating the forward and reverse shock radii. 
In this phase the amount of matter swept up by the shock fronts is more than the ejecta mass. During ST the energy radiated by the hot gas ($E_{\rm rad}$), that is enclosed by the shock fronts, is negligible. However, when these losses become significant the blastwave enters the snow plough phase. This transition roughly occurs at $T_{\rm rad}$ when
\beq
E_{\rm rad} \approx  \Gamma ~  n_{H}^2 ~ \frac{4}{3}\pi r_s^3 ~ T_{\rm rad} \sim E_k,
\label{eq:trad}
\eeq
where $E_k$ is the kinetic energy of the ejected material. $\Gamma$ represents the cooling function. $n_{H}$ is the density of the hot gas, and it is assumed to be equal to the electron density ($n_e$) on that medium, i.e., $n_{H} = n_e$.
In case the shock expands in a constant density medium (s=0) then $T_{\rm rad}$ is given by
\beq
T_{\rm rad} \sim 5 \times 10^{4} ~ \bigg(\frac{E_k}{10^{51} {\rm erg}}\bigg)^{0.22} ~ ~ \bigg(\frac{n_{\rm {ISM}}}{1 ~cm^{-3}} \bigg)^{-0.55} ~ ~ \rm{yrs}
\label{eq:T_ST}
\eeq
\citep{draine11}. If we assume that $T_{\rm ST}$ is the duration of ST phase then $T_{\rm ST} \sim T_{\rm rad}$.


\section {Dispersion Measure and Rotation Measure}
\label{sec:DM_RM}

In both the CC and merger cases the shocks are supersonic in nature and have speeds of the order of few tens $\times 10^3$ km/sec.
Therefore, as the forward shock moves into the CSM it sweeps up material and compresses the matter into a thin shell. Similarly, the reverse shock squeezes the matter in the ejecta. For a strong shock the material is compressed by a factor 4 in the shocked region. The temperature of this shocked shell is also very high, $\gsim 10^6$ K \citep{chevalier82,vink11, kundu19} . Therefore, the region between the reverse and the 
forward shock contains matter that is ionised and has high density. If the magnetar/NS that formed due to a merger and/or CC is a source of FRB then the emitted pulses will be dispersed as they pass through the shocked shell. The DM due to this shell is
\beq
 {\rm DM}_{\rm sh} = \int_{r_{\rm rev}}^{r_s} n_e (r) dr = \int_{r_{\rm rev}}^{r_c} n_e^{\rm rev} (r) dr + \int_{r_{c}}^{r_s} n_e^{s} (r) dr, 
 \label{eq:DM_sh}
 \eeq 
where $n_e^{\rm rev} (r)$ and $n_e^{s} (r)$ are the electron density of the shocked ejecta and the shocked CSM, respectively. 
Here $n_e^{s} (r)= 4 A r^{-s}/\mu m_p$, where $s$= 0 (2) for the merger (CC) channel. Therefore, in the free expansion phase the DM due to the shocked CSM can be written as
\beq
{\rm DM}_{\rm sh,csm}^{\rm FE} = \frac{4 ~ A~ \phi_{\rm csm}}{\mu~ m_p (1-s)} ~ D^{\frac{1-s}{n-s}}  ~ t^\frac{(n-3)(1-s)}{(n-s)},
\label{eq:DM_FE_fw}
\eeq
where $\phi_{\rm csm} = (\alpha^{1-s} -1 )$. The contribution due to the shocked ejecta is 
 \beq
{\rm DM}_{\rm sh,ej}^{\rm FE} = \frac{4 ~ A~\phi_{\rm ej}}{\mu~ m_p (1-s)} ~ D^{\frac{1-s}{n-s}} ~ t^\frac{(n-3)(1-s)}{(n-s)},
\label{eq:DM_FE_rev}
\eeq
where $\phi_{\rm ej} = ~\frac{(n-3)(n-4)}{(3-s)(4-s)} ~ (1- \beta^{1-s} )$, and  $n_e^{\rm rev} (r) = n_e^{s} (r) \frac{(n-3)(n-4)}{(3-s)(4-s)}$ is the density of the shocked ejecta. We assume here that $n_e^{\rm rev} (r)$ is similar to the reverse shock density that can be obtained from the thin shell approximation of the shocked shell \citep[cf.,][]{chevalier82,kundu19}. Therefore, the DM in the ST phase is given by
\beq
{\rm DM}_{\rm sh,csm/ej}^{\rm ST} = \frac{4 ~ A~ \phi_{\rm csm/ej}}{\mu~ m_p (1-s)} ~ D^{\frac{1-s}{n-s}} ~ T_{\rm FE}^{\frac{(n-5)(3-s)(1-s)}{(n-s)(5-s)}} ~ t^{\frac{2(1-s)}{5-s}}.
\label{eq:DM_ST_fw_rev}
\eeq

\par
The shocked shells are ideal places for the amplification of magnetic fields \citep{bykov13,caprioli14}.
As a result, these shells are often bright at radio wavelengths due to the emission of synchrotron radiation from charged particles.
In shocks, usually, a fraction of the post shock energy  ($\epsilon_{\rm B}$) is channeled into the magnetic field. Therefore, the energy density of the magnetic field in the post shock region is given by
\beq
\frac{\rm B^2}{8 \pi} = \epsilon_{\rm B} ~ u_{\rm th}. 
\label{eq:B}
\eeq
Here ${\rm B}$ represents the magnetic field strength and $u_{\rm th} = 9\rho_{\rm csm}v_s^2/8$ with $v_s$ being the velocity of the forward shock. The modeling of radio emission from the SN shocks suggests that around 10\% of the kinetic energy converts into magnetic field  \citep[e.g., see][]{chevalier06, soderberg12}. Therefore, in this work we assume that $\epsilon_{\rm B} = 0.1$. When a radio pulse passes though this magnetised material it may change the plane of polarisation of the wave. The magnitude of this change is given by the rotation measure (${\rm RM}$), and is written as
\beq
{\rm RM} = K \int_{r_{\rm rev}}^{r_s} n_e (r) B_{\parallel}\,dr 
\label{eq:RM}
\eeq  
where $K = \frac{e^3}{2 \pi m_e^2 c^4}$. $e$ is the electric charge, $m_e$ is the mass of an electron and $c$ represents the speed of light in vacuum. ${\rm B}_{\parallel}$ is the component of magnetic field along the line of sight. We assume that ${\rm B} =  {\rm B}_{\parallel}$.  Therefore, in the free expansion phase RM due to the shocked CSM and ejecta are
\beq
\begin{split}
{\rm RM}_{\rm sh,csm/ej}^{\rm FE}  = 4K ~  \frac{({9 \pi \epsilon_{\rm B} A^3})^{1/2}}{\mu m_p} \psi_{\rm csm/ej} ~ \frac{2 \alpha  (n-3)}{(2-3s)(n-s)}  \\ D^{\frac{(4-3s)}{2(n-s)}} ~ t^{\frac{2(n-6)+s(11-3n)}{2(n-s)}},
\end{split}
\label{eq:RM_FE_fw_rev}
\eeq
where $ \psi_{\rm csm} = \left(\alpha^{(1-3s/2)} -1 \right)$ and  $ \psi_{\rm ej} =\displaystyle{ \frac{(n-3)(n-4)}{(3-s)(4-s)}} ~ \left(1 - \beta^{(1-3s/2)} \right)$. It is considered here that the magnetic field strength is same across the shocked shell. The RM contribution in the ST phase is
\beq
\begin{split}
{\rm RM}_{\rm sh,csm/ej}^{\rm ST}  = 4K ~  \frac{({9 \pi \epsilon_{\rm B} A^3})^{1/2}}{\mu m_p} \psi_{\rm csm/ej} ~ \frac{4 \alpha}{(5-s)(2-3s)}  \\ D^{\frac{(4-3s)}{2(n-s)}} ~ T_{\rm FE}^{\frac{(n-5)(3-s)(4-3s)}{2(n-s)(5-s)}} ~ t^{\frac{2s+1}{s-5}}.
\end{split}
\label{eq:RM_ST_fw_rev}
\eeq

\subsection{Merger of two WDs}
\label{sebsec:Merger_WDs}
 
\begin{table}
 \caption{ Ejecta mass and $E_k$ of three WD merging models.
 $a$ from \citet{dessart07}, \citet{fryer99}
 $b$ from \citet{metzger09} 
 $c$ from \citet{dessart06} }
 \label{tab1}
 \begin{tabular}{@{}lcc}
  \hline
  Model & ejecta mass ($\msun$) & $E_k$ (erg) \\
  \hline
  WDcaseA$^{a}$ & $\sim$ 0.2 & $\sim 5 \times 10^{50}$  \\
  WDcaseB$^{b}$ & $\sim$ 0.01 & $\sim 5 \times 10^{49}$  \\
  WDcaseC$^{c}$ & $\sim$ 0.001 & $\sim 5 \times 10^{48}$  \\
  \hline
 \end{tabular}
 \label{table:WDmodel}
\end{table}

\begin{table}
 \caption{ $T_{\rm{FE}}$ for three merger models and different values of $n_{\rm ISM}$ and $n$.}
 \begin{tabular}{@{}lccc}
  \hline
  Model   & $n_{\rm ISM}$ ($\ccc$) & $n$ & $T_{\rm{FE}}$ (yr)\\
  \hline
  WDcaseA  & 50    &  10  &  11.5       \\
           &       &  6   &  29.4     \\
           & 10    &  10  &  19.7       \\
           &       &  6   &  50.3     \\
           & 1     &  10  &  42.5       \\
           &       &  6   &  108    \\
           & 0.1   &  10  &  91.6       \\
           &       &  6   &  233   \\
           & 0.01  &  10  &  197    \\
           &       &  6   &  503      \\
\hline
  WDcaseB  & 50    &  10  &  3.7        \\
           &       &  6   &  9.5      \\
           & 10    &  10  &  6.3      \\
           &       &  6   &  16.2     \\
           & 1     &  10  &  13.7       \\
           &       &  6   &  35      \\
           & 0.1   &  10  &  29.5     \\
           &       &  6   &  75      \\
           & 0.01  &  10  &  63.5    \\
           &       &  6   &  162     \\
\hline
  WDcaseC  & 50    &  10  &  2.1     \\
           &       &  6   &  4.4     \\
           & 10    &  10  &  3       \\
           &       &  6   &  7.5     \\
           & 1     &  10  &  6.3     \\
           &       &  6   &  16.2    \\
           & 0.1   &  10  &  13.7    \\
           &       &  6   &  35     \\
           & 0.01  &  10  &  29.5   \\
           &       &  6   &  75    \\
  \hline
 \end{tabular}
 \label{table:WDmodel_2}
\end{table}
When a degenerate ONe core approaches the Chandrasekhar's mass, its collapse is triggered by electron capture on $^{24}$Mg and $^{20}$Ne before a thermonuclear runaway, leading to a SN event, can occur \citep[e.g.][]{Nomoto87, Woosley15}. Such a collapse can happen when two massive WDs, where at least one of them is an ONe WD, coalesce \citep{dessart06, dessart07}. As the angular momentum is conserved during  collapse, a fast differentially rotating WD is produced with some residual mass left behind in an accretion disk. The fast rotation allows the critical mass for electron-capture collapse to substantially surpass the Chandrasekhar's limit \citep{Yoon05,dessart06}. For instance, the simulations of \citet{dessart06} find that if the initial ratio between the rotational energy to the gravitational binding energy of the nascent white dwarf is 0.0833, its mass can reach 1.92\msun. Thus, a merger induced collapse can lead to the formation of NSs that are generally more massive and spin much more rapidly than those produced by CC SNe. It is still not clear what the rotationally delayed merger-induced collapse of massive white dwarfs is, but the work of \citet{Tornambe13} indicates that the redistribution or loss of angular momentum can lead to delays of up to a few millions of years while \citet{IlkovSoker2012} suggest delay times that are even longer ($10^6-10^{10}$ years). The consequence of such long delays may be the reason why radio and X-ray observations of the media surrounding SNe Ia are so tenuous \citep{chomiuk16, margutti12, kundu17} and with a density profile constant with radius.

\par 
The theoretical work of \citet{dessart06} has shown that the electron capture induced collapse causes the ejection of a few $\times 10^{-3}$\msun of material with a typical velocity of $0.1\,c$ and energies $\le 10\times^{50}$\,erg just after the NS/magnetar is born. Although small ejecta masses ($<10^{-2}$) have also been found by other investigators \citep[e.g][]{Yoon05}, the simulations of \citet{fryer99} have yielded ejection masses considerably larger, around $0.2$\msun. It is not clear why the computed mass of ejecta differs so much in these works but it is likely to be caused by the different equation of states employed and numerical codes. The MHD simulations of \citet{dessart07} have revealed that if the collapse is aided by a magnetic field, the stellar core is significantly spun down with the rotational energy converted into magnetic energy. This causes the generation of strong magnetically driven winds leading to the ejection of about 0.1 $\msun$ of material. Interestingly, \citet{dessart07} find that the explosion energy immediately after bounce is also much higher (about $10^{51}$\,erg) than in models in which the presence of fields is neglected. They also find that despite the extraction of angular momentum by magnetic fields, the proto-neutron star would still be rotating at $\sim1$\,ms period so that a magnetic field of a few $10^{15}$\,G is likely to be generated in the process, following the $\alpha-\Omega$ mechanism \citep{Duncan92, Thompson93, Ferrario2015, Wick2014}. 

\par
In summary, since there is no strong consensus about the amount of mass ejected during a double white dwarf merger event, in this paper we present three models, summarised in Table \ref{table:WDmodel},  within the $10^{-3} - 0.3$\msun range.

\par
\begin{figure*}
\centering
 \includegraphics[width=8.5cm,origin=c]{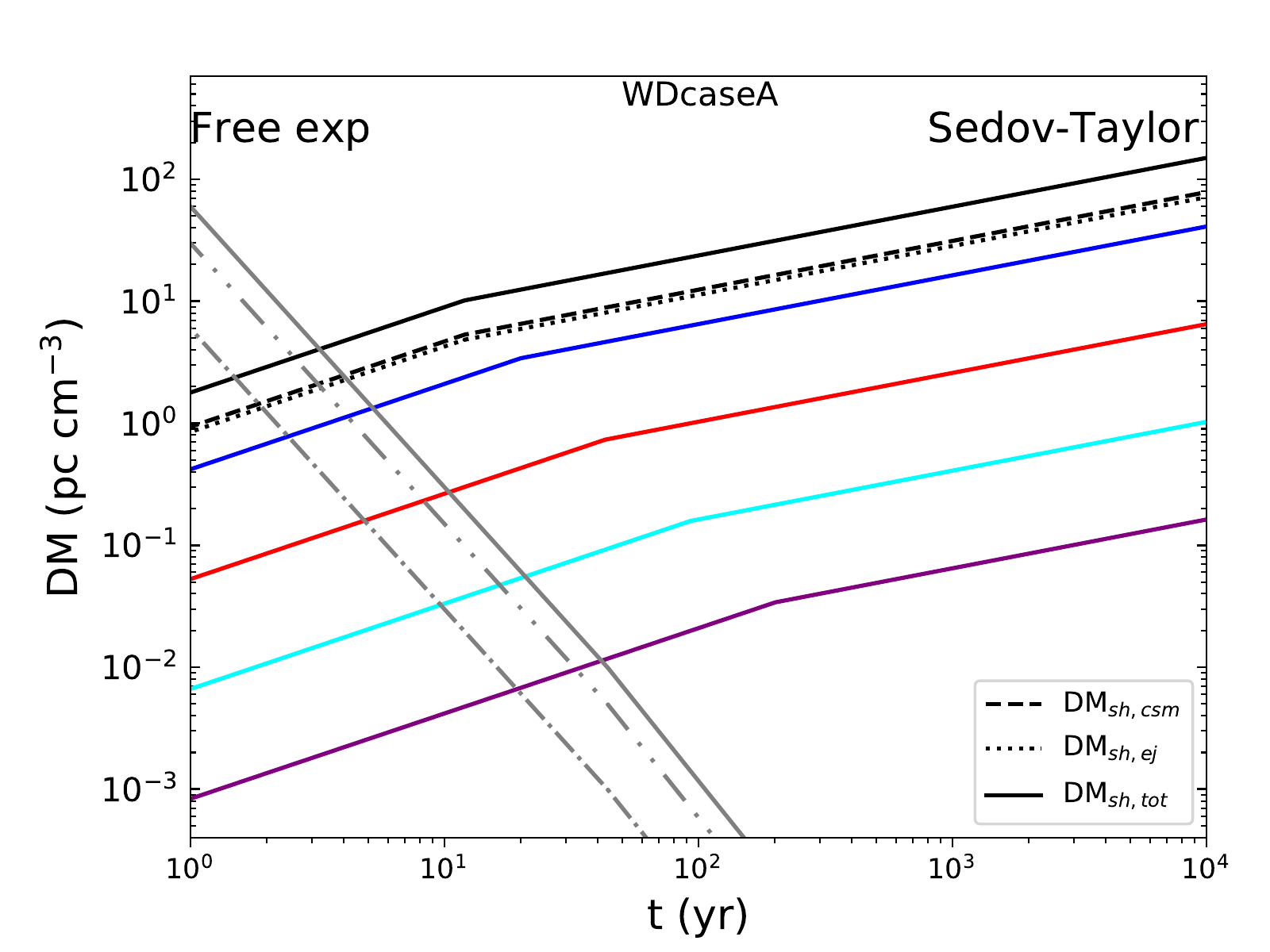}
 \includegraphics[width=8.5cm,origin=c]{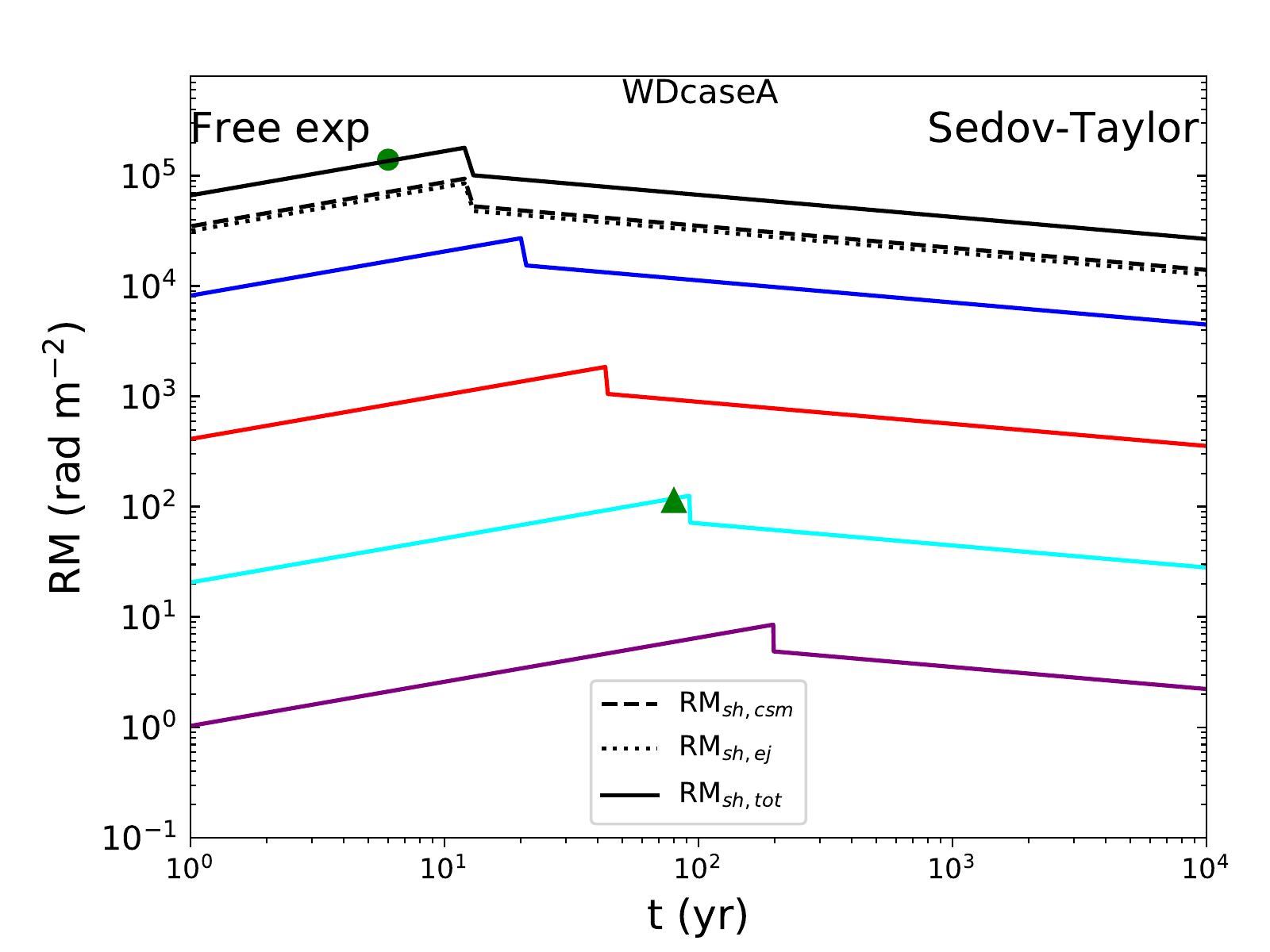}
 \includegraphics[width=8.5cm,origin=c]{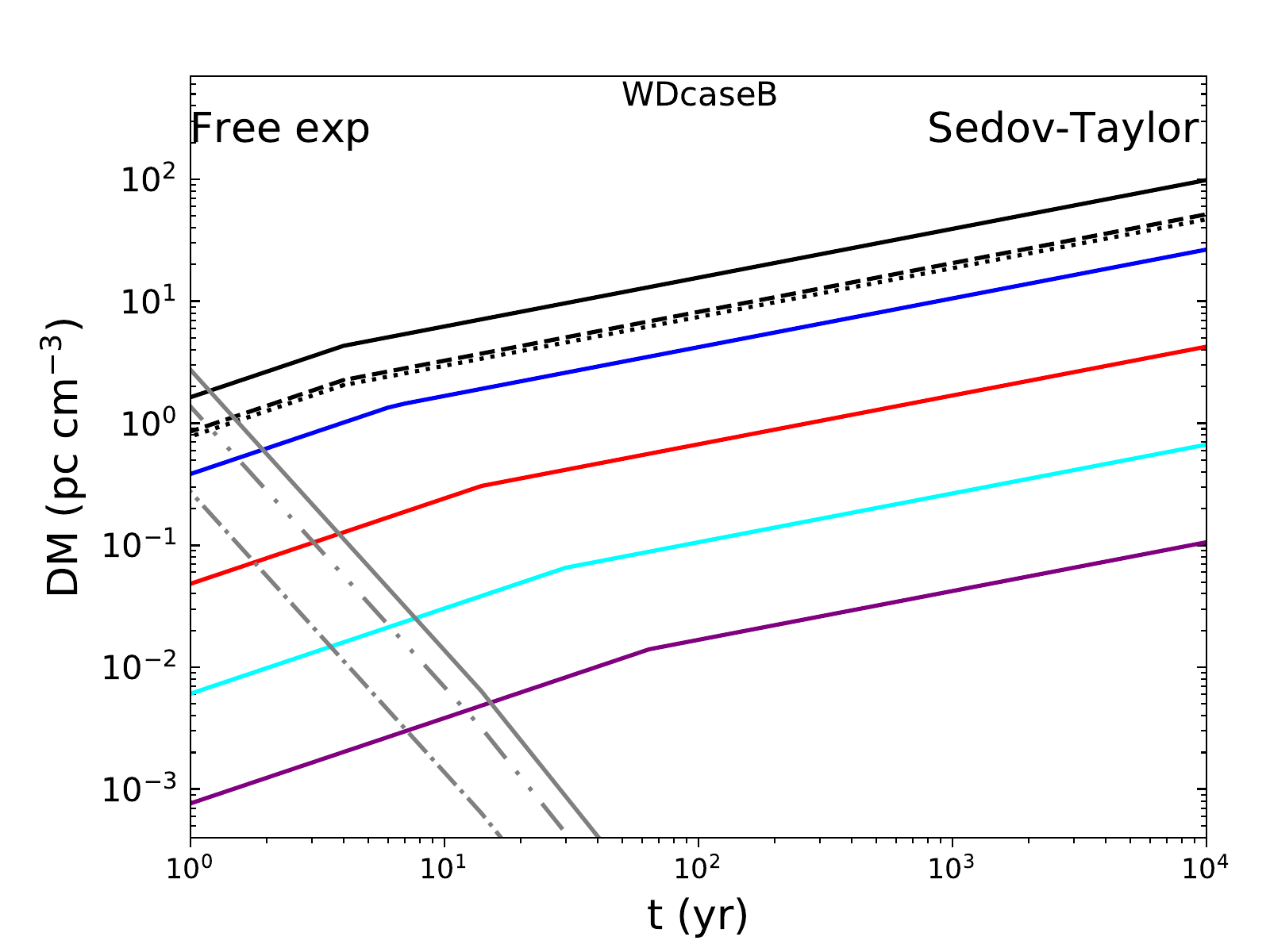}
 \includegraphics[width=8.5cm,origin=c]{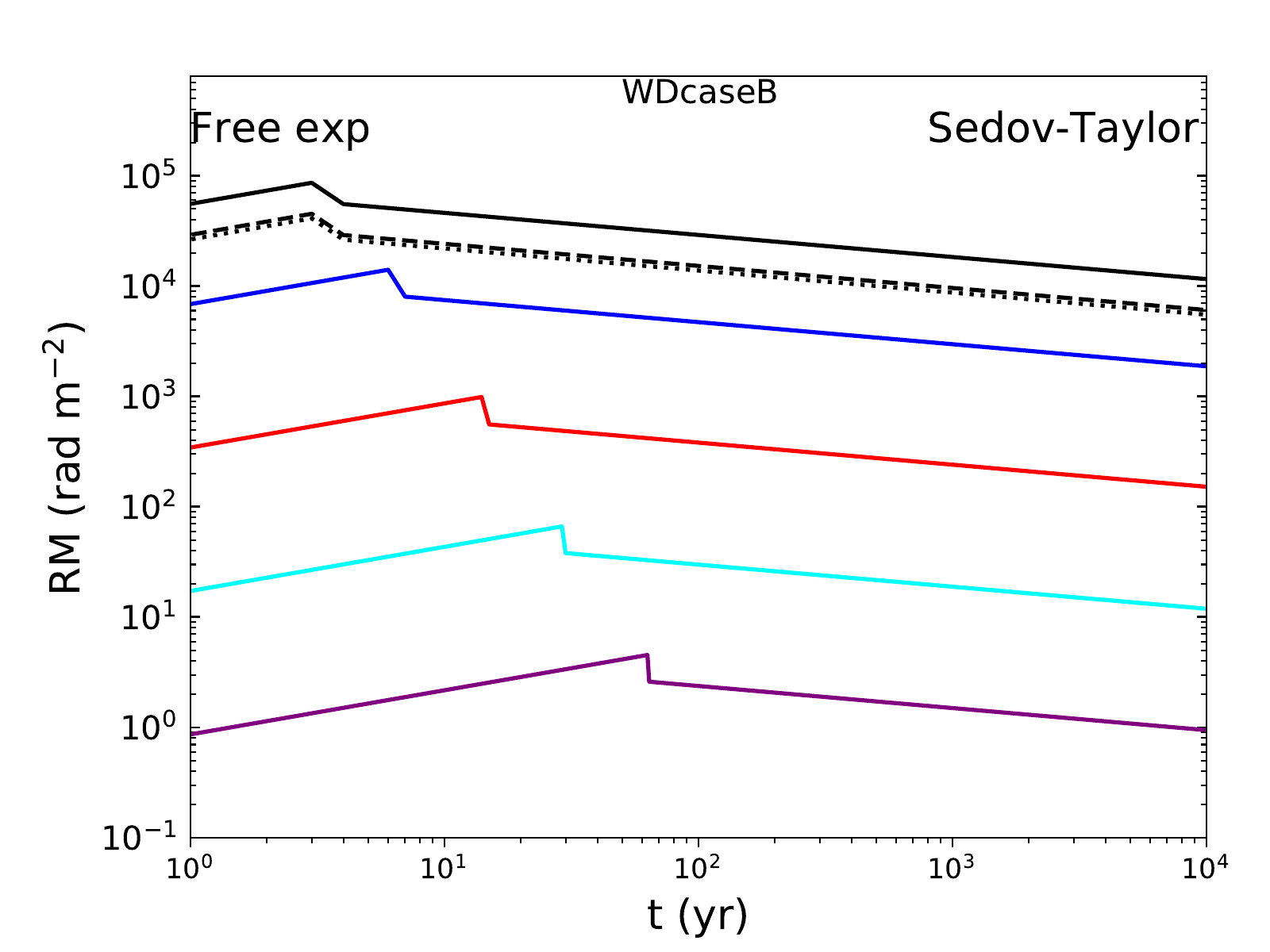}
 \includegraphics[width=8.5cm,origin=c]{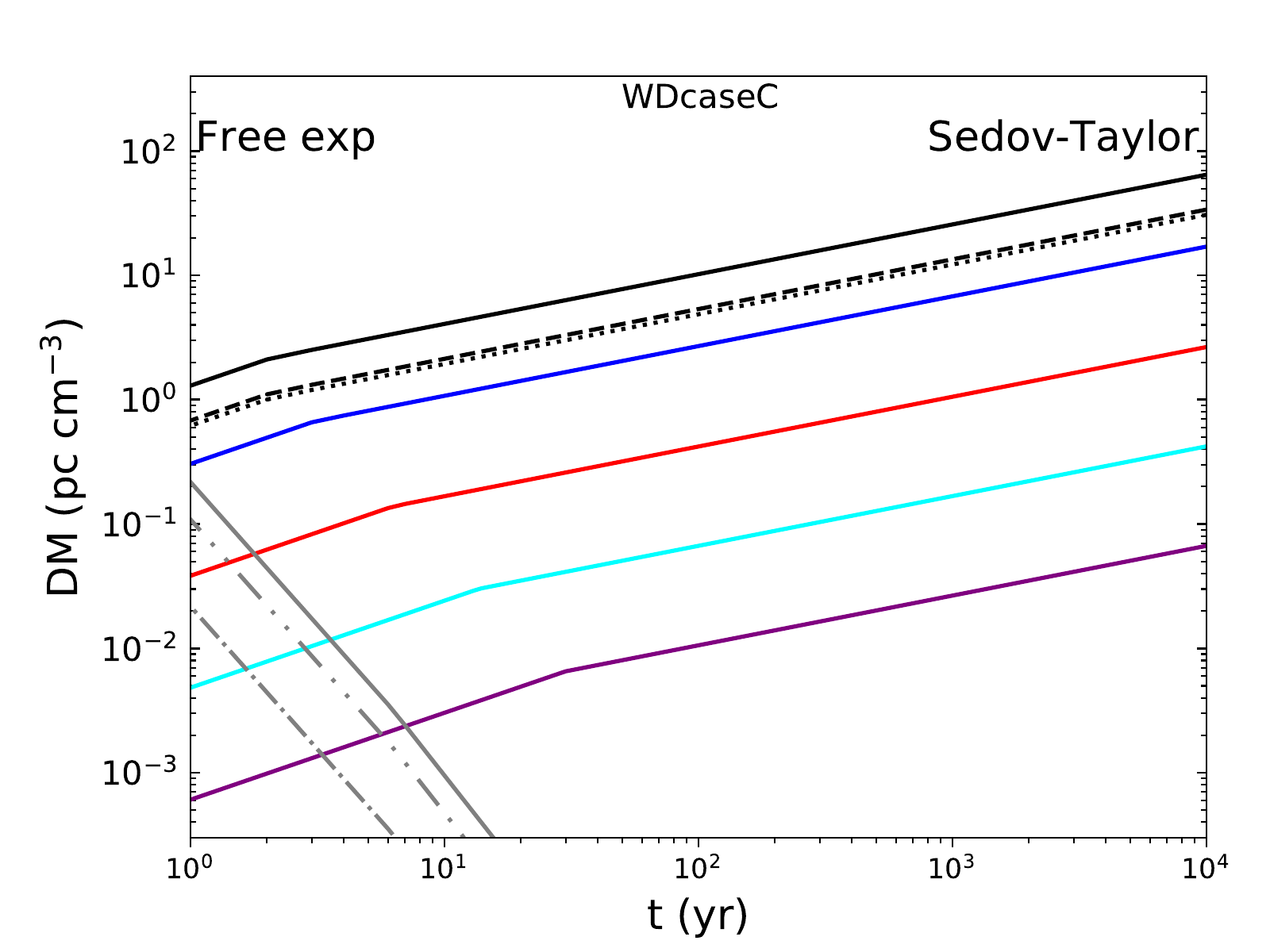}
 \includegraphics[width=8.5cm,origin=c]{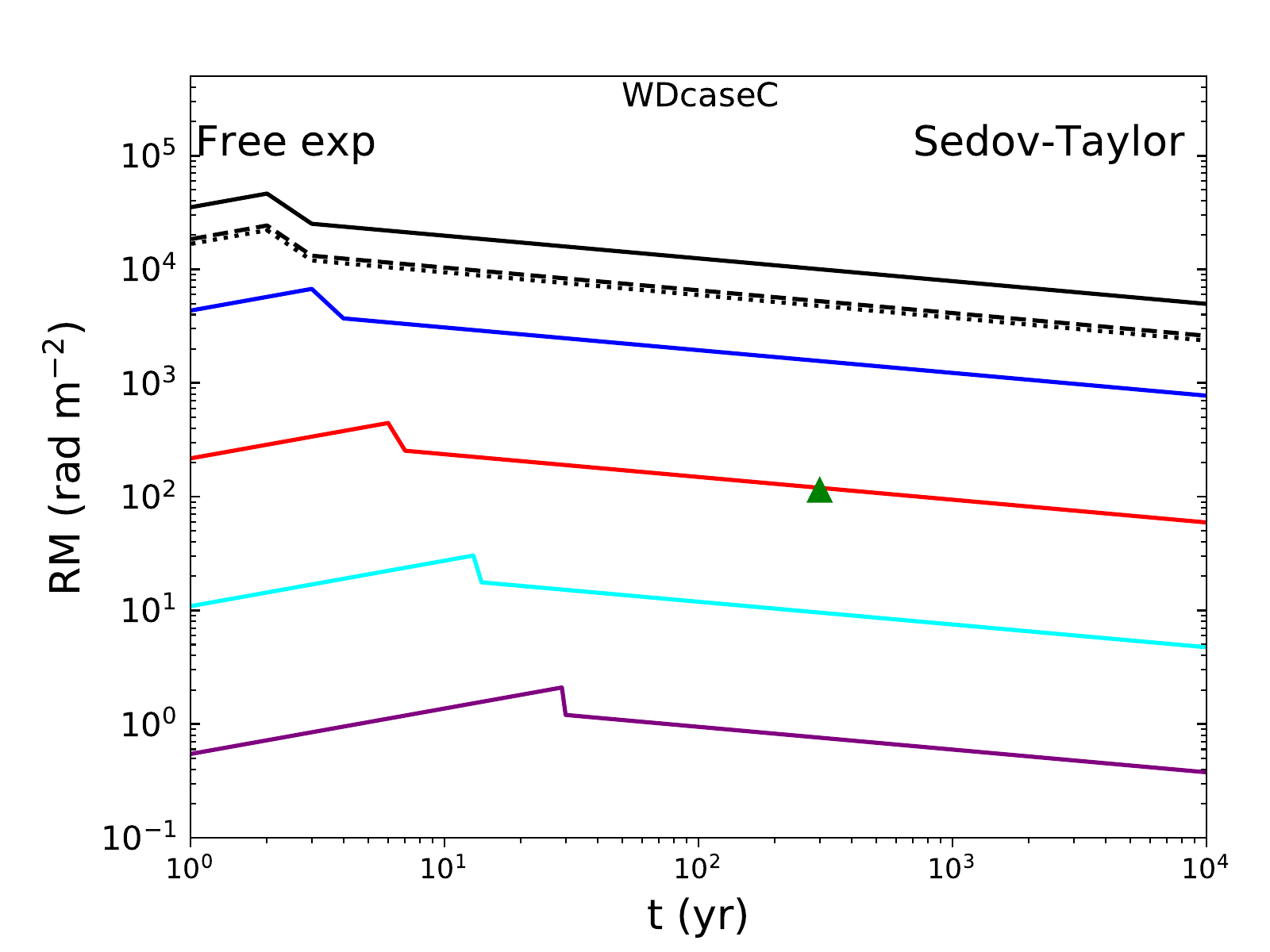}
 \caption{Evolution of DM (left panel) and RM (right panel) in the free expansion and ST phases for the three merger models considered here. The ejecta with a power-law index of $n = 10$ plough through a constant density medium having  $n_{\rm ISM} =$ 50 $\ccc$ (black line), 10 $\ccc$ (blue line),  1 $\ccc$ (red line), 0.1 $\ccc$ (cyan line) and 0.01 $\ccc$ (purple line). The dashed and the dotted lines illustrate the contributions due to the shocked CSM and ejecta, respectively. The solid, dash double-dotted and dash-dotted gray lines show the DM expected from the ejecta when the ionisation fractions are 100\%, 50\% and 10\%, respectively. This material is not supposed to be magnetised in general, therefore, no contribution to RM from the ejecta. The kink in the curves represents the transition from the free expansion to ST phase.  In the left panels, the RM of FRB\,121102 and FRB\,180916.J0158+65 are displayed with green circles and triangles, respectively. For WDcaseB a $n_{\rm ISM}$ between 1 $\ccc$ and 0.1 $\ccc$ can account for the observed RM of FRB\,180916.J0158+65. In case of FRB\,121102 and WDcaseA the corresponding DM contributions from the shocked shell and ionised ejecta are 6.3 $\rm pc ~ cm^{-3}$, and 0.1 $\rm pc ~ cm^{-3}$ (10\% ionised), 0.5 $\rm pc ~ cm^{-3}$ (50\% ionised) and 1 $\rm pc ~ cm^{-3}$ (100\% ionised)  respectively.
 }
 \label{fig:merger_DM_RM_n10}
\end{figure*}

\begin{figure*}
\centering
 \includegraphics[width=8.5cm,origin=c]{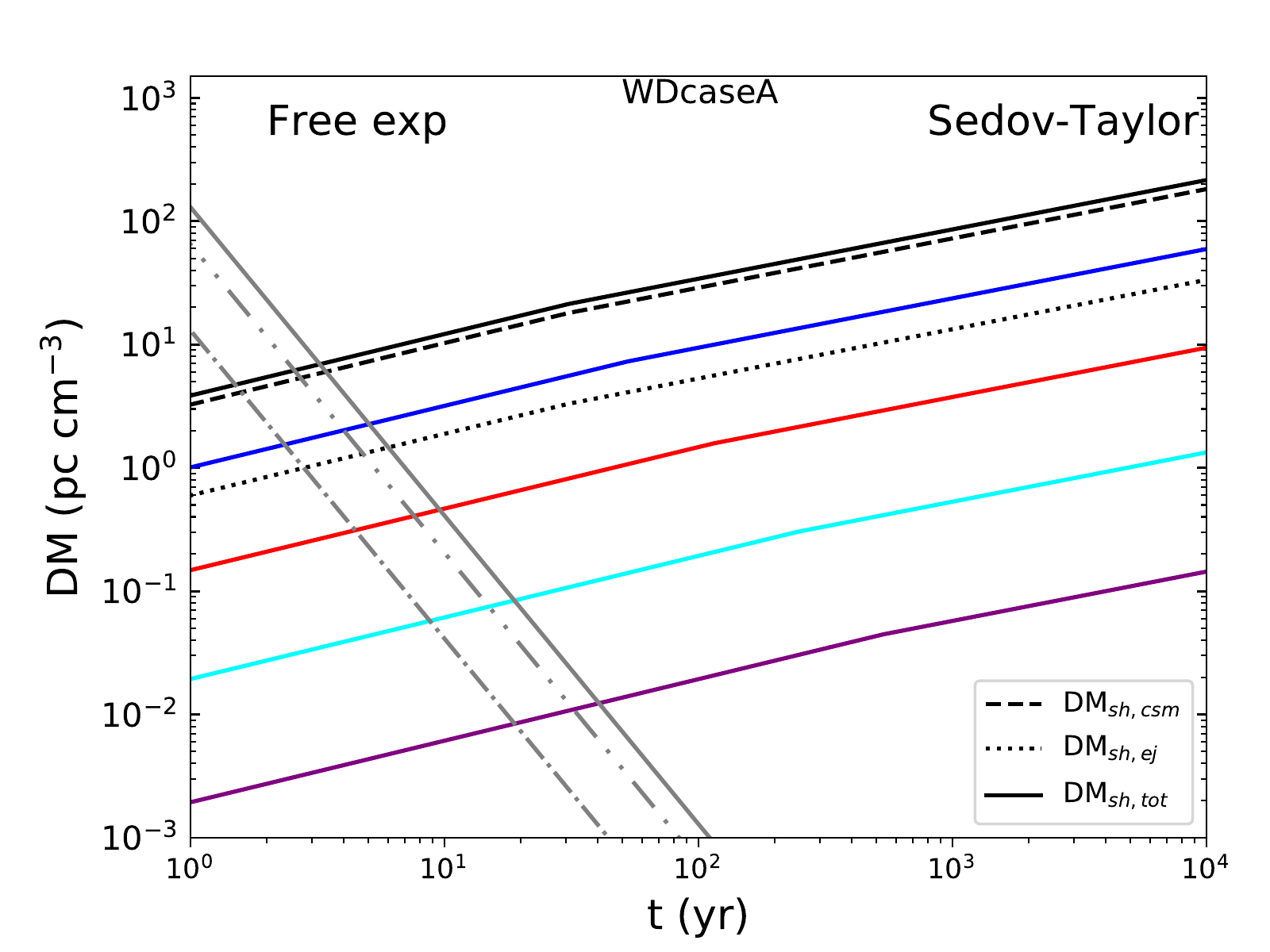}
 \includegraphics[width=8.5cm,origin=c]{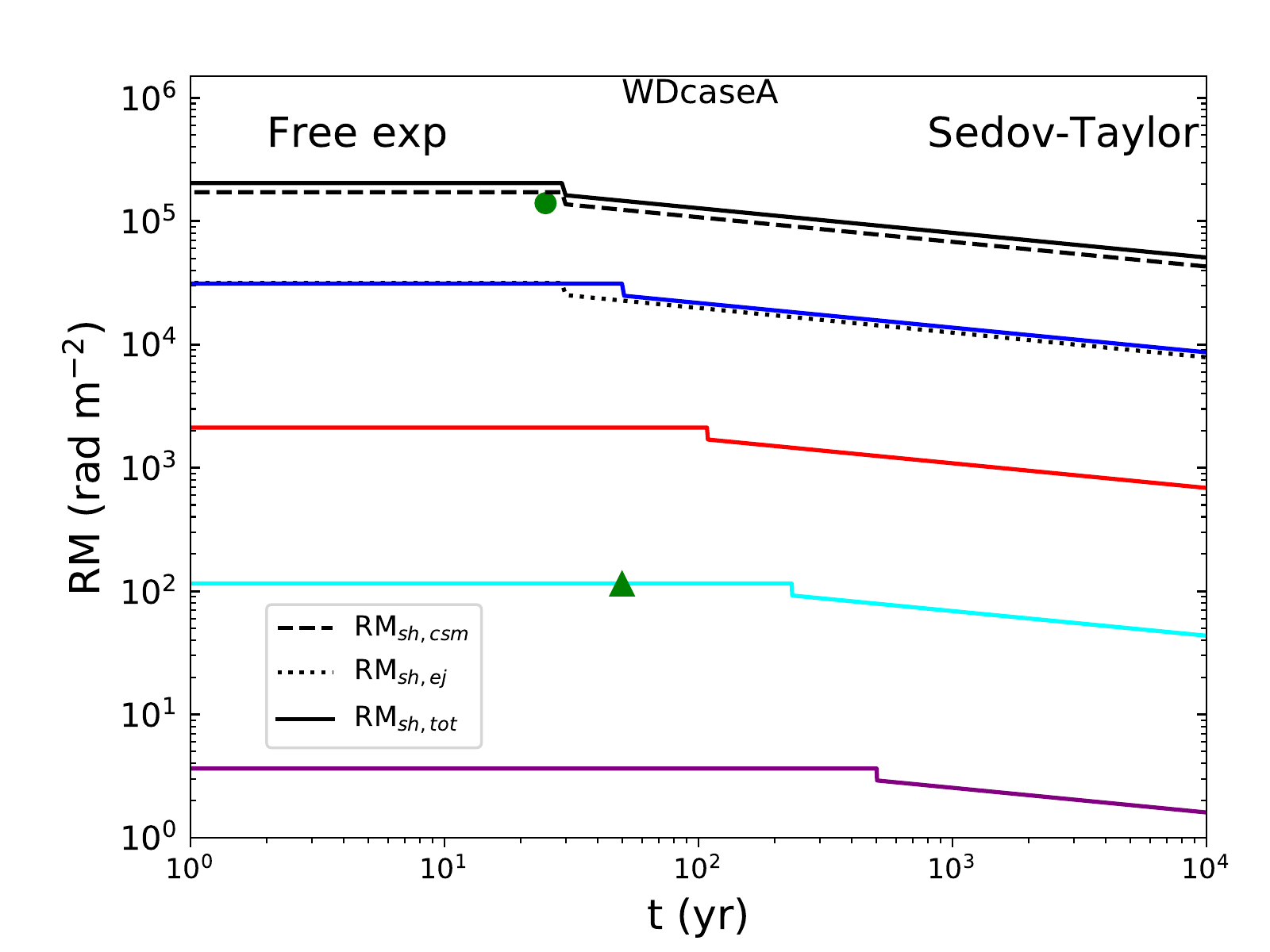}
 \includegraphics[width=8.5cm,origin=c]{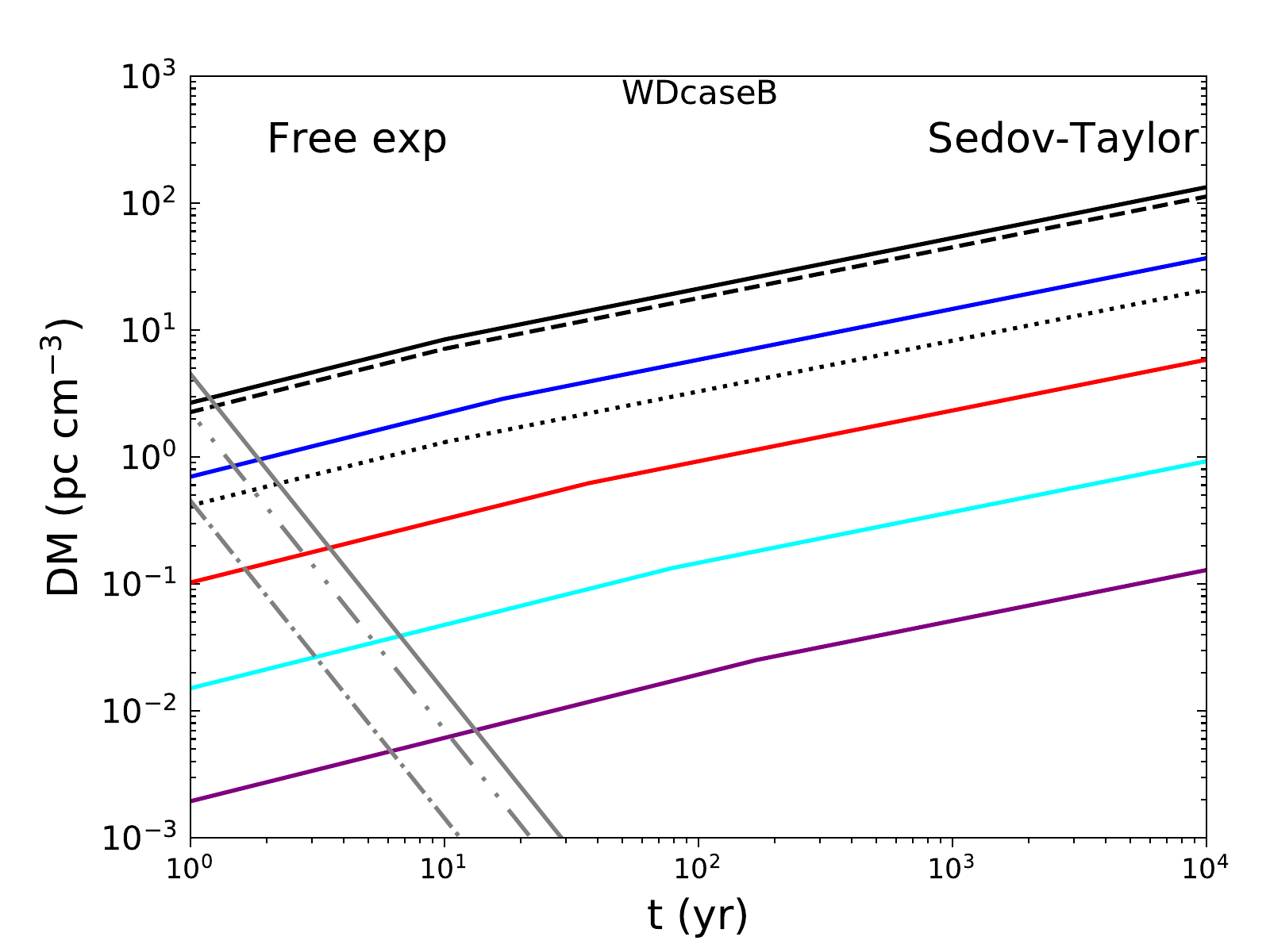}
 \includegraphics[width=8.5cm,origin=c]{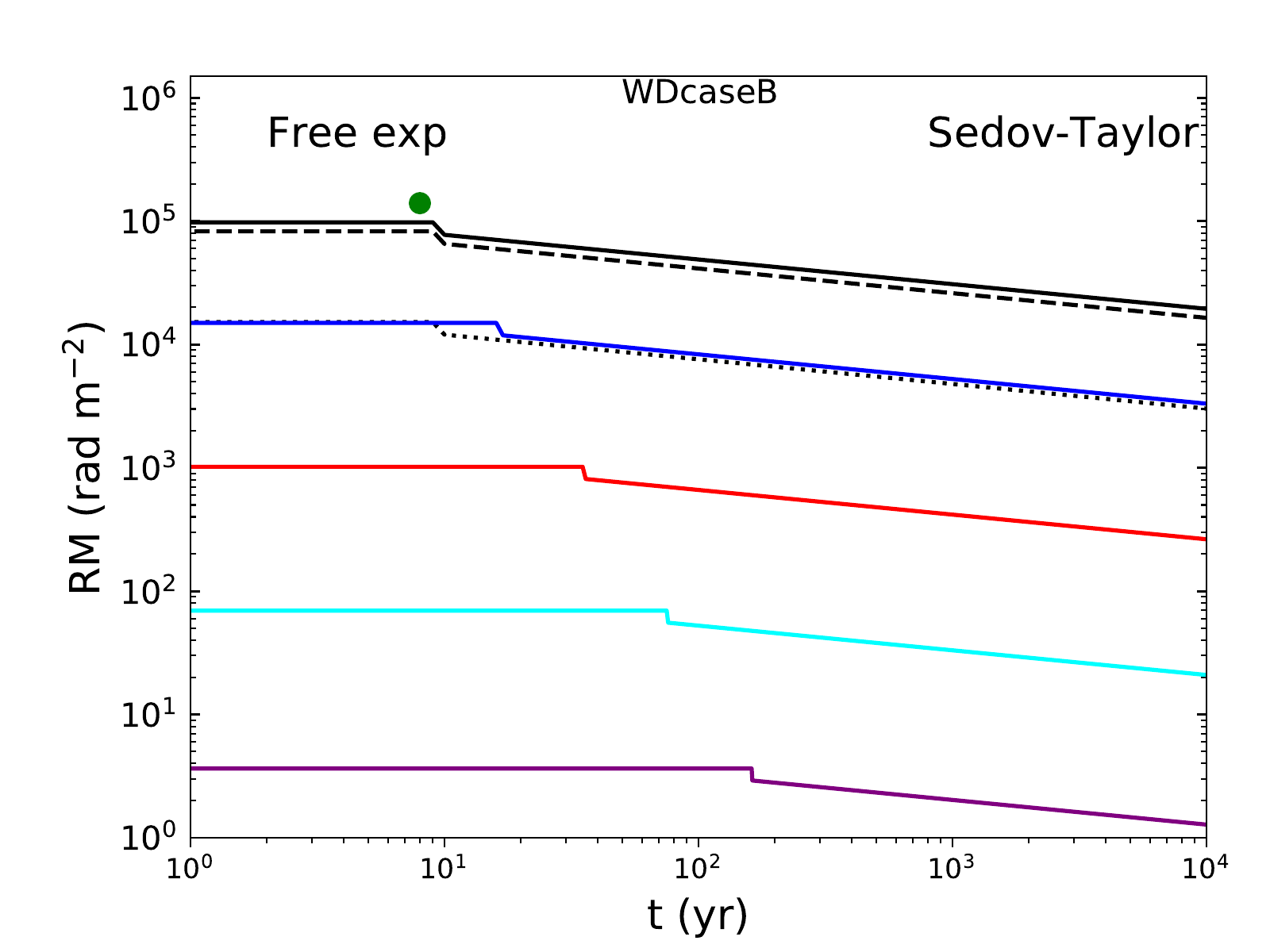}
 \includegraphics[width=8.5cm,origin=c]{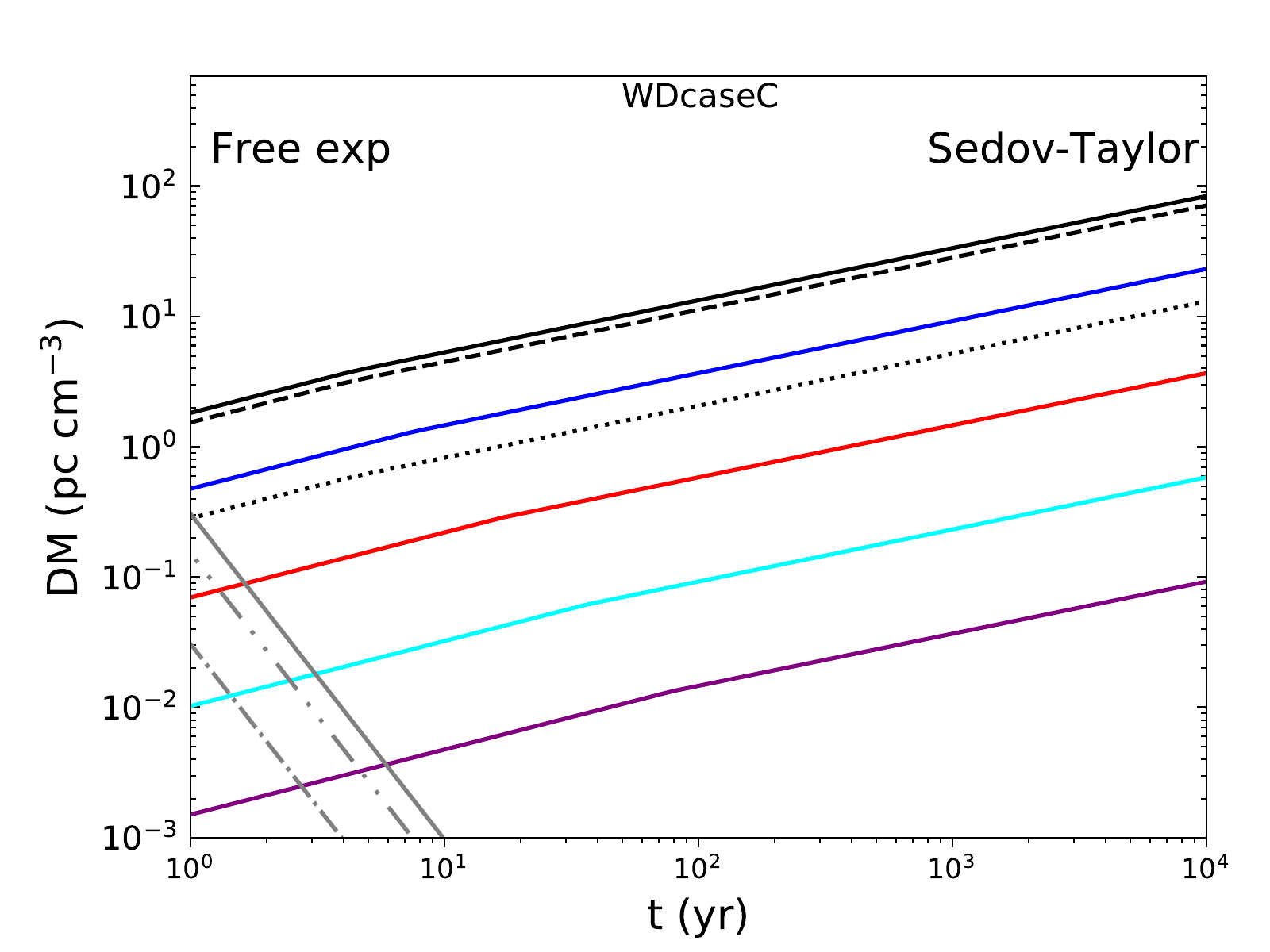}
 \includegraphics[width=8.5cm,origin=c]{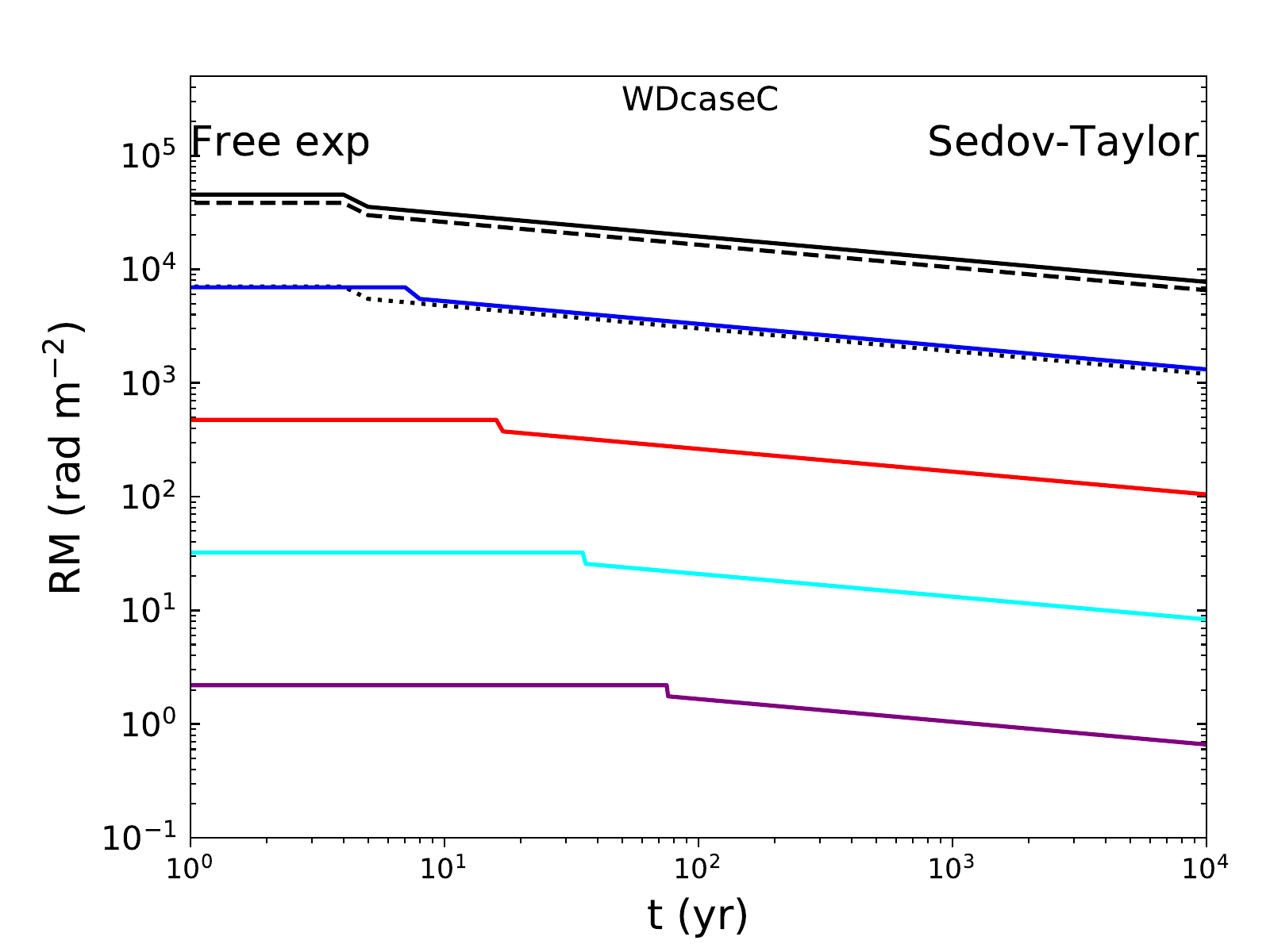}
 \caption{Similar to figure \ref{fig:merger_DM_RM_n10} except the ejecta in this case have a power-law index (n) of 6. In the upper and middle right panels, the RM of FRB\,121102 are displayed with green circles as the estimated RM from these two models are close to that has been observed for this repeater FRB. In case of WDcaseA the corresponding DM contributions to this FRB from the shocked shell and ionised ejecta are 21 $\rm pc ~ cm^{-3}$, and 0.002 $\rm pc ~ cm^{-3}$ (10\% ionised), 0.012 $\rm pc ~ cm^{-3}$ (50\% ionised) and 0.024 $\rm pc ~ cm^{-3}$ (100\% ionised)  respectively . The RM of FRB\,180916.J0158+65 is displayed with a green triangle in the upper right panel. As RM is constant in the free expansion phase this FRB is placed in the midway of cyan line in the plot. For other models, WDcaseB and WDcaseC, a $n_{\rm ISM}$ between 1 $\ccc$ and 0.1 $\ccc$ can account for the observed RM of FRB\,180916.J0158+65.
 }
 \label{fig:merger_DM_RM_n6}
\end{figure*}

 
As mentioned earlier, the ejecta have a flat inner part density distribution while beyond $v_{\rm brk} \sim$10000 $\kms$ \citep{pakmor12,kundu17} the density structure is given by a power-law. The mass and kinetic energy ($E_{\rm k}$) of the ejected material considered in this work are tabulated in Table \ref{table:WDmodel}. The evolution of DM and RM for $n = 10$ are shown in Figure \ref{fig:merger_DM_RM_n10} for the three models. While the dashed and the dotted lines illustrate the contributions from the shocked CSM and ejecta, respectively, the solid lines (black, blue, red, cyan and purple) depict the total contribution (${\rm DM}_{\rm sh,tot}$ and ${\rm RM}_{\rm sh,tot}$), i.e., ${\rm DM}_{\rm sh,tot} = {\rm DM}_{\rm sh,ej} + {\rm DM}_{\rm sh,CSM}$ and ${\rm RM}_{\rm sh,tot} = {\rm RM}_{\rm sh,ej} + {\rm RM}_{\rm sh,CSM}$. Here the evolution are shown for $n_{\rm ISM} =$ 50 $\ccc$, 10 $\ccc$,  1 $\ccc$, 0.1 $\ccc$ and 0.01 $\ccc$ with black, blue, red, cyan and purple lines, respectively. The DM and RM evolution for $n = 6$ are shown in Figure \ref{fig:merger_DM_RM_n6} for the different models.
The ionisation factors depend on the object generated by the collapse. In the case of a rapidly spinning ($1$\,ms) magnetar, which is the most likely outcome of a double white dwarf merger, its spin-down energy is expected to fully ionize the ejecta \citep[e.g.,][] {Margalit2019}. Therefore, in this work we do not make an assumption on the type of object that may be generated by the collapse, rather, we present calculations that will make predictions for different ionising factors. Our work will allow to identify the nature of the ionising source once more data of FRBs, and thus constraints on possible models, become available. Hence we have considered ionisation fractions of 100\%, 50\% and 10\%. The solid, dash- double dotted and dash-dotted gray lines show the DM expected from the ejecta when the ionisation fractions are 100 \%, 50\% and 10\%, respectively. It should be noted that for $n = 6$ the ejecta are in the free expansion phase for a longer duration in comparison to the $n=10$ situation. For different merger models and CSM densities $T_{FE}$ are tabulated in Table \ref{table:WDmodel_2}. The DM and RM exhibit a rapid changeover at $T_{FE}$ because we did not consider a smooth ejecta profile across $v_{\rm brk}$, for which one would get a smooth transition of DM and RM across $T_{FE}$. Computing the evolution of the shocked shell when the slope of the ejecta changes smoothly requires one to perform hydrodynamical simulations of the ejecta-CSM interaction, which is beyond the scope of this paper. The absolute values of RM and DM across the transition can be up to $\sim$ 10\% different in the case of a smooth profile. For the three models $T_{ST}$ is $\sim$ $10^{4}$ yrs (see eq. \ref{eq:T_ST}). Therefore, the evaluations of DM and RM are shown for around 10,000 yrs.

\subsection{Core collapse explosion }
\label{sebsec:CC}

\begin{table}
 \caption{ $T_{\rm{FE}}$ for different values $\mdot$ when $v_w = 10~ \kms$.}
 \label{tab1}
 \begin{tabular}{@{}lcc}
  \hline
  $\mdot$ ($\msunyr$) &  $n$ & $T_{\rm{FE}}$ (yr)  \\
  \hline
 $1 \times 10^{-4}$ &   10     &   12      \\
                    &    6     &   131      \\
\hline                    
$1 \times 10^{-5}$  &   10     &  119       \\
                    &    6     &  1314       \\  
\hline                    
$1 \times 10^{-6}$  &   10     &   1186      \\
                    &    6     &  13137       \\  
  \hline
 \end{tabular}
 \label{table:CC_TFE}
\end{table}

 Massive stars are usually born in molecular clouds and end their lives as CC supernova. The stellar wind and photoionising radiation from the progenitor stars play an important role in removing the cloud and interstellar matter from the vicinity of the star \citep{weaver77,dwarkadas05}. In case of B0-O4 stars these processes can clear a region up to a radius $\sim$ 30 pc \citep{mckee84}. For the supernova remnant (SNR) VRO 42.05.01 it is found by \citet{arias19} that the wind has blown up a region of radius of around 10 pc. Thus, in these cases, the SN propagates in a wind like medium for a few couple of thousand years before it starts to interact with a wind-blown bubble. 


\par
\begin{figure*}
\centering
 \includegraphics[width=8.5cm,origin=c]{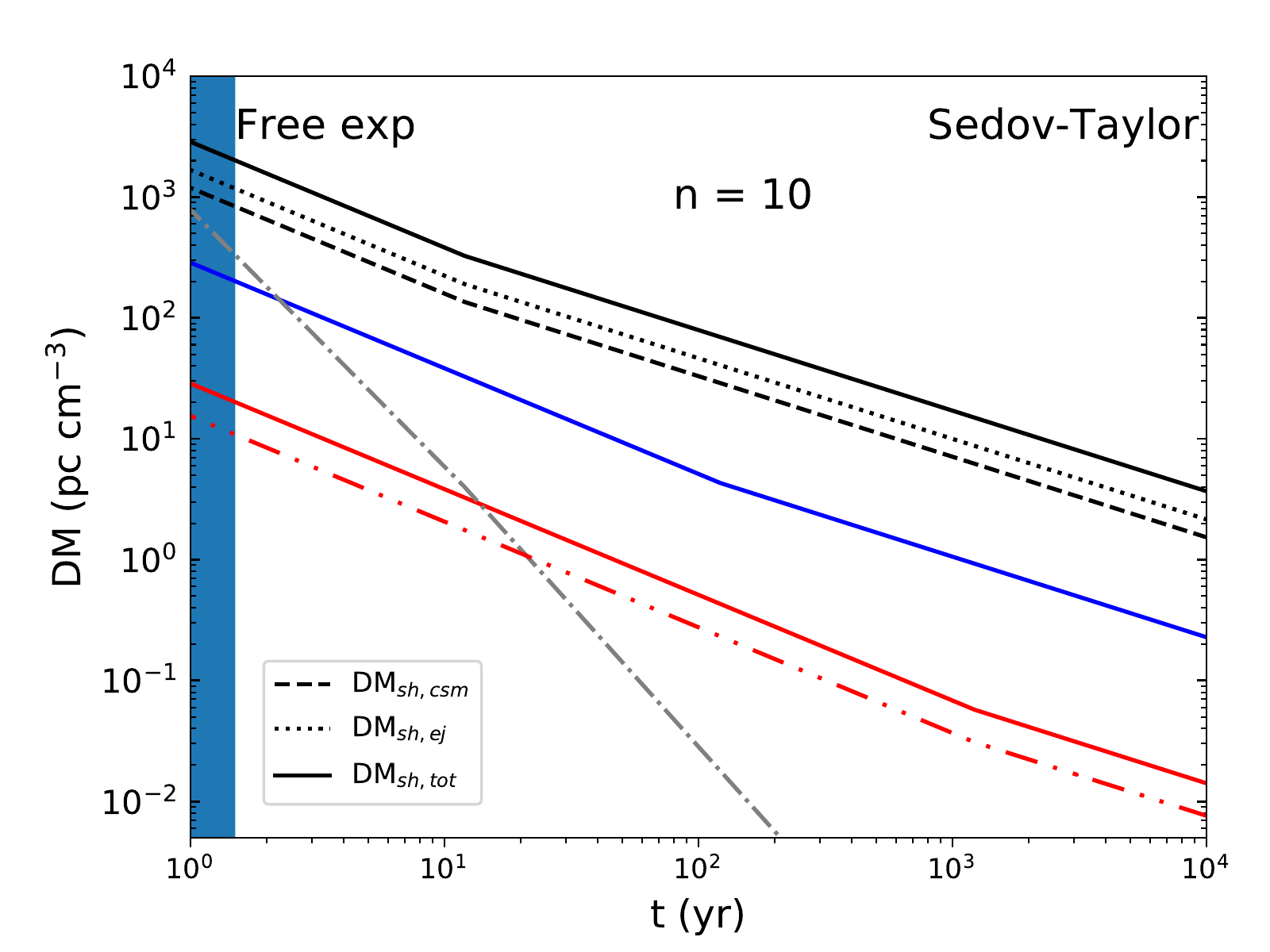}
 \includegraphics[width=8.5cm,origin=c]{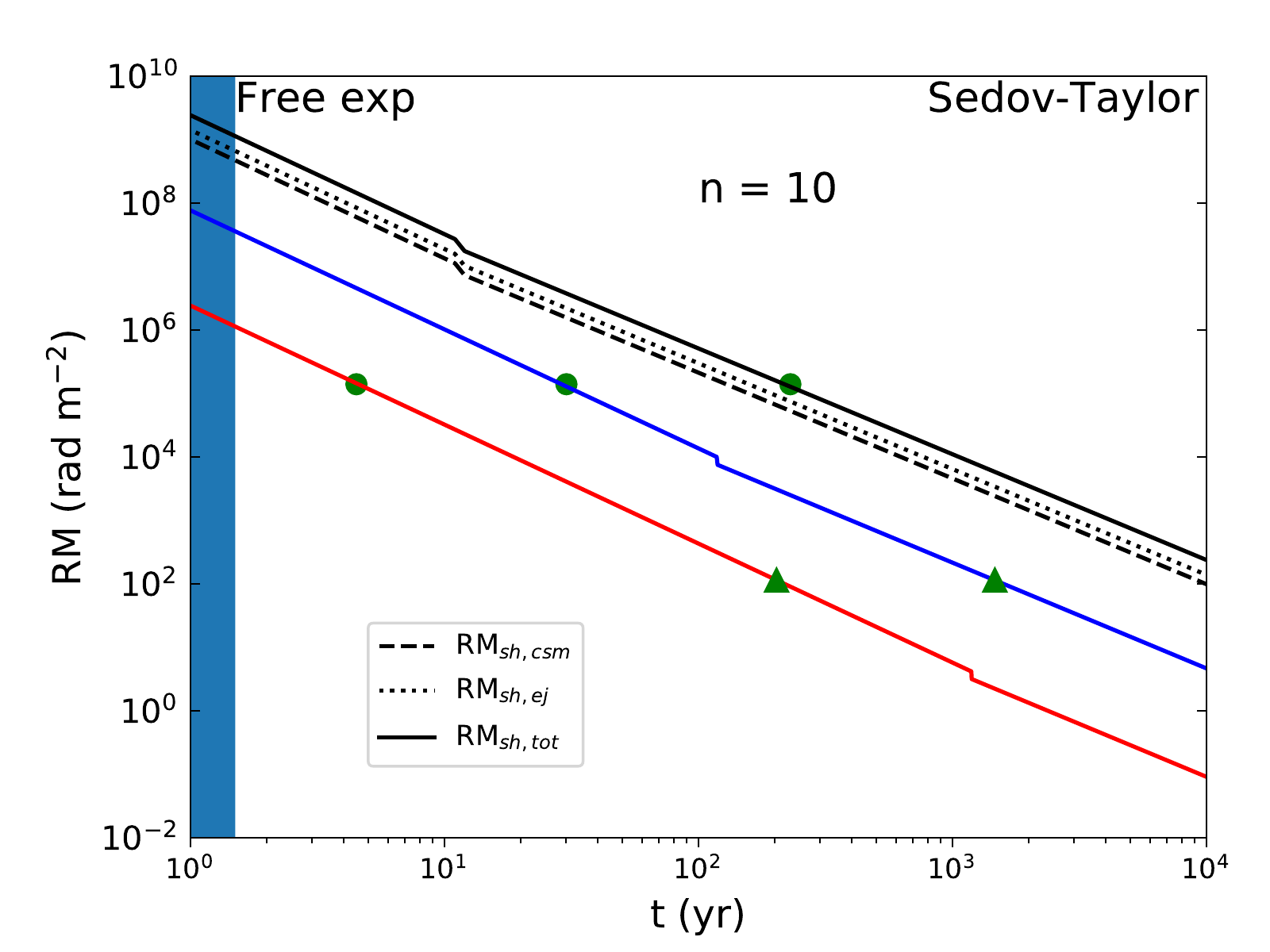}
 \includegraphics[width=8.5cm,origin=c]{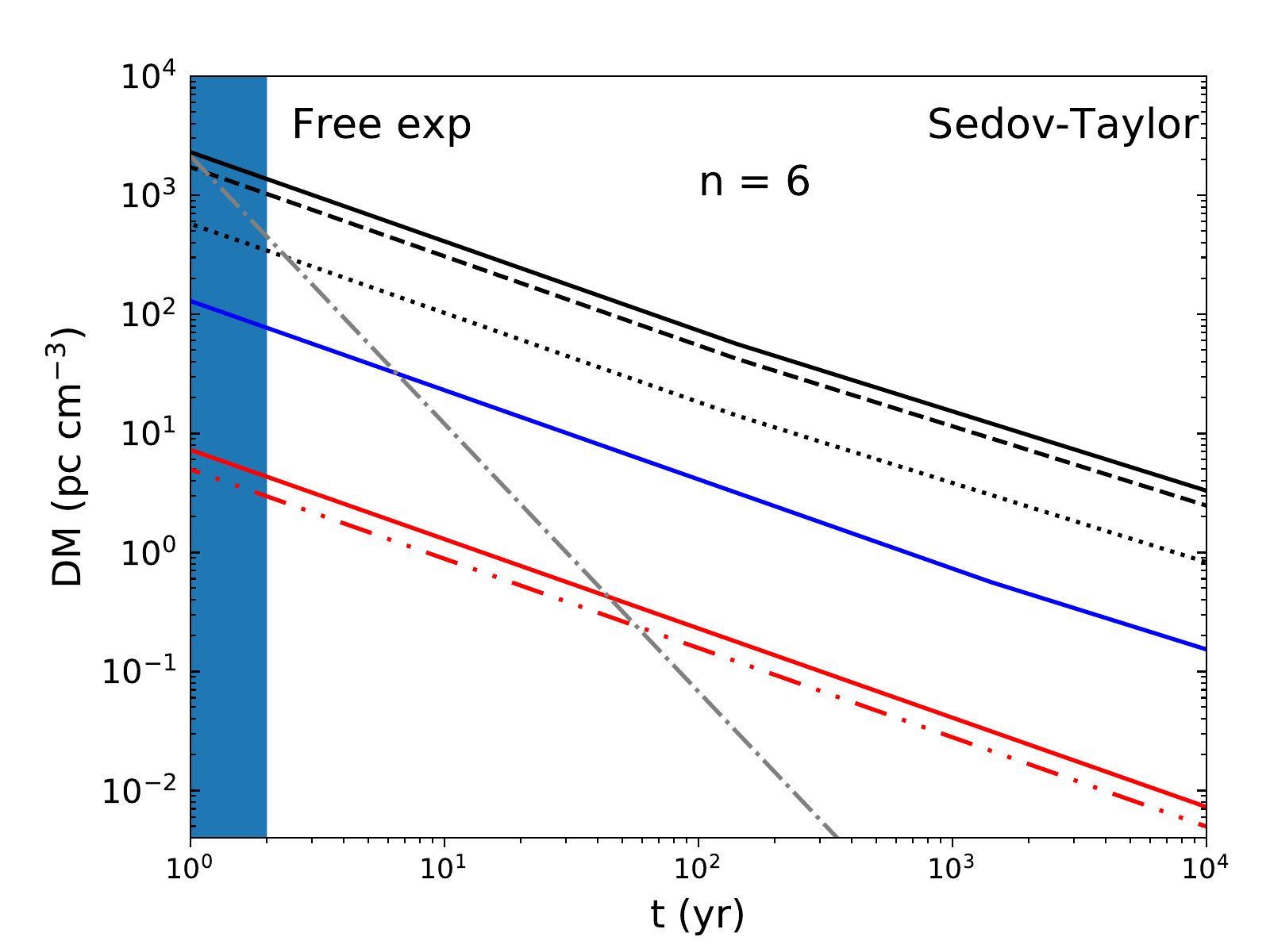}
 \includegraphics[width=8.5cm,origin=c]{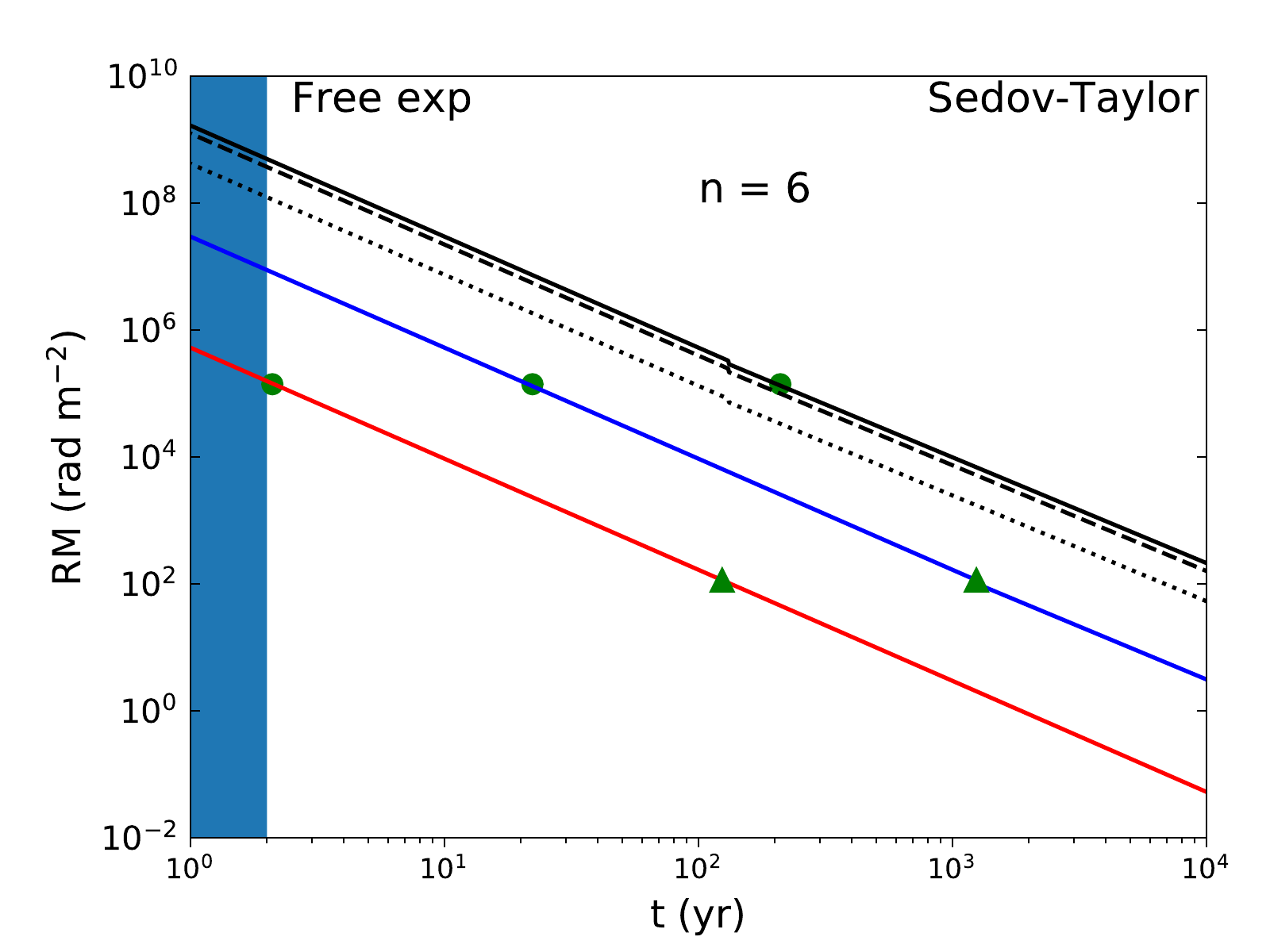}
 \caption{Evolution of DM (left panel) and RM (right panel) in the free expansion and ST phases for the CC channel. Here the ejecta having a mass of 5 $\msun$ and $n= 10$ (upper panels) or $n= 6$ (lower panels) interact with a wind medium characterized by $\mdot = 1\times 10^{-4}$ $\msunyr$ (black curve),  $1\times 10^{-5}$  $\msunyr$ (blue curve) and $1\times 10^{-6}$  $\msunyr$ (red curve) for a wind velocity $v_w $ of 10 $\kms$. The dashed and the dotted lines illustrate the contributions due to the shocked CSM and ejecta, respectively. The dash-dotted gray line indicates the DM expected from the ejecta when this material is ionised by a fraction of about 3\%. Because of the presence of the free electrons the ejecta remain optically thick for around 1.5 yrs and 2 yrs when $n = 10$ and 6, respectively. This is shown with the royal blue shaded region in the plots. The red dash-double dotted line represents the DM from the unshocked ambient medium when the CSM, characterised by  $ \mdot/v_w = 1\times 10^{-6}~ \msunyr/10 ~\kms$, is assumed to be ionised up to infinity. 
 The kink in the curves represents the transition from the free expansion to ST phase. As for the merger scenario, the green circles and triangles display the RM of FRB\,121102 and FRB\,180916.J0158+65, respectively.  In case of FRB\,121102 the corresponding DM contributions, for $\mdot = 1\times 10^{-6}~ \msunyr$, $1\times 10^{-5}~ \msunyr$ and $1\times 10^{-4}~ \msunyr$, from the shocked shell are 4.3 $\rm pc ~ cm^{-3}$, 13 $\rm pc ~ cm^{-3}$ and 43 $\rm pc ~ cm^{-3}$ for $n=6$ (8 $\rm pc ~ cm^{-3}$, 14.5 $\rm pc ~ cm^{-3}$ and 45.4 $\rm pc ~ cm^{-3}$ for $n=10$), respectively. The DM inputs from the ionised ejecta are 448 $\rm pc ~ cm^{-3}$, 2 $\rm pc ~ cm^{-3}$ and 0.01 $\rm pc ~ cm^{-3}$ for $n=6$ (33 $\rm pc ~ cm^{-3}$, 0.5 $\rm pc ~ cm^{-3}$ and 0.007 $\rm pc ~ cm^{-3}$ for $n=10$).
 }
 \label{fig:CC_DM_RM}
\end{figure*}

 
For the CC explosion we consider a SN ejecta of 5 $\msun$ with $v_{\rm brk} \sim 5000\,\kms$ and $E_{\rm k} = 1 \EE{51}$\,erg. For two different values of $n$, 10 and 6, the DM and RM due to the shocked shell in the free expansion and ST phases are shown in Figure \ref{fig:CC_DM_RM}. The black, blue and red lines demonstrate the cases when the ejecta interact with an ambient medium characterized by $\mdot = 1\times 10^{-4}$ \msunyr,   $1\times 10^{-5}$  \msunyr and $1\times 10^{-6}$  \msunyr, respectively, for a wind velocity $v_w $ of 10 \kms. $T_{FE}$ for the three values of $\mdot/v_w$ considered here are tabulated in Table \ref{table:CC_TFE}. In case of core collapse SN\,1993J it is estimated by \citet{chevalier16} that the ionisation fraction in the inner ejecta is around 3\%. Considering a similar ionisation fraction for the ejecta we estimate the DM, which is shown with dash-dotted grey lines in figure \ref{fig:CC_DM_RM}.

\par 
 In case of $\mdot = 1\times 10^{-4}$ \msunyr the shock velocity at $10^4$ yrs, for both values of $n$ (10 and 6), is around 800 $\kms$ from our calculations. Therefore, the temperature of the shocked gas is $3\mu m_p v_s^2/(16 k_B) \sim 10^7$ K,  where $k_B$ is the Boltzmann constant and $\mu \simeq 1$. The cooling function, $\Gamma$, at this temperature is around $10^{-23}$ erg cm$^{3}$ sec$^{-1}$ for a solar-abundance plasma in collisional ionisation equilibrium \citep{draine11}. According to our model the radius of the SNR is 10 pc and $n_{H} \sim 5$ cm$^{-3}$ at this age. This implies that even after $10^4$ yrs of evolution the energy radiated by the shocked shell, $E_{\rm rad}$, is much less than $E_k$ (see eq. \ref{eq:trad}) . Thus, the shocks have not yet entered in the radiative phase. For $\mdot = 1\times 10^{-5}$  \msunyr and $1\times 10^{-6}$  \msunyr the shocks decelerate at a slower rate compare to $\mdot = 1\times 10^{-4}$ \msunyr, as the density of the CSM is low. As a result, the shock temperature is higher than $10^7$ K, which means that with a $\Gamma \sim 10^{-23}$ erg cm$^{3}$ sec$^{-1}$ \citep{draine11} and $n_{H} < 1$ cm$^{-3}$ there will be less amount of cooling . 
Therefore, the shocks are not in the radiative phase even though the radius of the shock is around 25 pc. 
Beyond an age of around $10^4$ yrs, when the shock radius is $\sim 20$ pc, the SN may start to interact with a wind bubble.
In that case the onset of the snow-plough phase is better estimated by numerical simulations, which is beyond the scope of this paper.To make it easy to compare the CC scenario with the merger channel, in this paper, we have shown the evolution of the DM and RM up to 10,000 yrs.  



\section{Discussion} 
\label{sec:dis}
For the merger channel the DM may be dominated by the ionised ejecta in the first few years to couple of decades (see solid, dash double-dotted and dash-dotted gray lines in figures  \ref{fig:merger_DM_RM_n10} and \ref{fig:merger_DM_RM_n6}) depending on the density of the CSM. The same is true for CC unless the $\mdot/v_w > 1\times 10^{-4}~ \msunyr/10 ~\kms$ (see Figure \ref{fig:CC_DM_RM}). As the ejecta evolve over time, the density of this medium decreases, and, therefore, for homologous expansion the DM declines as $t^{-2}$, which implies a rapid drop of these quantities after the merger or CC. This material is not supposed to be magnetised in general, therefore, there will be no contribution to RM from the ejecta. Depending on the density of the ionized particles and temperature of this medium the radio pulses may suffer from free-free absorption. For a plasma with a temperature $T$ and electron density $n_e$ the free-free absorption coefficient at a frequency $\nu$ is given by 
\beq
\alpha_{\nu}^{\rm ff} = \frac{4 e^6}{3 c m_e k_B } ~\bigg(\frac{2 \pi}{3 k_B m_e}\bigg)^{1/2} ~ T^{-3/2} ~ z^2 ~ n_e ~ n_i ~ \nu^{-2} ~ \bar{g}_{\rm ff}
\label{eq:freefree}
\eeq
\citep{ribicki79}, where $\bar{g}_{\rm ff}$ is the velocity average gaunt factor, $z$ is the atomic number of the ion, which has a density $n_i$. 
With a temperature of $10^5$ K \citep{Margalit2018,Margalit2019} the ejecta for the three merger models, with $n = 10$ and 6, become transparent to 1 GHz within a month after the formation of the NS/magnetar.  In case of CC the temperature of the ejecta is considered to be $10^4$ K from the calculations done by \citet{chevalier16} for SN\,1993J around 500 days after the explosion. The 1 GHz radiation, for $n=10$ and 6, will then trap in this medium up to around 1.5 years and 2 years, respectively. This is shown with the royal blue shaded region in figure \ref{fig:CC_DM_RM}. 




\par
Apart from the ejecta, another source of free-free absorption could be the shocked shell. The temperature of this shell in the free expansion and ST phases is usually $\gsim 10^{6}$ K \citep{vink11,chevalier82,kundu19}. As a result, the radio signals will not suffer from free-free absorption while passing through this medium at any point during the evolution. However, depending on the density of the ambient medium, the CSM may be opaque at 1 GHz signal at early times. For a CSM, having solar metallicity, and a temperature of $\sim 10^{5}$ K, which is found for SN 1993J \citep{claesI14}, the ambient medium becomes optically thin to free-free absorption around 1.5 yrs, for both values of $n$, after the collapse for a $\mdot/v_w = 1\times 10^{-4} \msunyr/10 \kms$.  For lower values of $\mdot/v_w$, that are considered here, the free-free absorption is not important even at very early epochs. This is also true for the merger of WDs where the shocked shell and the CSM are optically thin to the radiation right from the beginning unless $n_{\rm ISM} \gsim 10^{8}$ $\ccc$.

\par 
In these calculations we do not include the contribution made by the unshocked CSM. It is usually expected that a fraction of the CSM gets ionised because of the radiation from the shock breakout, shocked shell and hot ejecta. In the case of CC the DM from this medium is $\propto r_s^{-1}$, i.e., $\rm DM \propto t^{-(n-3)/(n-2)}$. If we assume that the radiation is able to ionise the ambient medium up to a large volume of radius $r_{max}$ such that $r_{max} \gg r_s$ then the DM from this medium will be comparable to that from the shocked shell. The red dash-double dotted line in the left panels of figure \ref{fig:CC_DM_RM} represents the DM from the unshocked ambient medium when the CSM, characterised by  $ \mdot/v_w = 1\times 10^{-6}~ \msunyr/10 ~\kms$, is assumed to be ionised up to infinity. However, $r_{max} \gg r_s$ is quite extreme and probably not attainable. Therefore, the DM contribution from the CSM ($\equiv \rm DM_{CSM}$) would be smaller than what has been shown here. In case of the merger model $\rm DM_{CSM} = \mu ~ (n_{\rm ISM}/1 \ccc) ~ (\Delta r({\rm pc})/1 {\rm pc}) ~ pc ~ \ccc$, where $\Delta r = r_{max} - r_s$ and $\mu \simeq$ 1.
Therefore, unless $\Delta r \gsim $ 1 pc and/or $n_{\rm ISM}$ is very large the $\rm DM_{CSM}$ would be negligible. As mentioned earlier, in the merger channel $n_{\rm ISM} \lsim 100 ~\ccc$. Initially, when $r_s \ll$ 1 pc, $\Delta r \sim 1$ pc is not feasible. After around $10^{4}$ yrs of evolution the $r_s\gg$ 1 pc. For instance, for both values of $n$ (6 and 10) and the three models considered here the $r_s \sim 10$ pc at an age of around $10^{4}$ yrs. Therefore, during ST phase for  $n_{\rm ISM} \sim 50~\ccc$ and $\Delta r \sim 1$ pc $\rm DM_{CSM} \simeq 50$ pc $\ccc$. 
The DM from the shocked shell at this age is $\gsim 100$ pc (see figures \ref{fig:merger_DM_RM_n10} and \ref{fig:merger_DM_RM_n6}). Furthermore, as the ambient medium is generally not expected to have a preferred orientation of the magnetic field there will be almost no contribution to RM from this medium. 

\par
For the CC channel it is expected that a small fraction of the inner ejecta would remain gravitationally bound to the central object \citep[for review, see][]{reynolds17}. According to the column density of free electrons this nebula would contribute to the total DM \citep[][]{Murase16}. 
Depending on the density and the temperature of the ionised gas this medium could be opaque to the 1 GHz radiation. The RM from the nebula is not important unless the nebula is highly magnetised. In the case of a merger induced collapse leading to the formation of a NS/magnetar, \citet{dessart07} find that these events are free of significant fallback.

\par 
The contribution to DM from the ionised ejecta decrease as they evolve.   As shown in Figures \ref{fig:merger_DM_RM_n10}, \ref{fig:merger_DM_RM_n6} and \ref{fig:CC_DM_RM} the DM from the shocked shells start to dominate the evolution of the total DM as early as few years since the formation of the NS/magnetar. When the density of the CSM is low it takes a few decades for the shocked shell to influence the DM evolution. The two cases discussed here demonstrate that the evolution of the ${\rm DM}_{\rm sh,tot}$ and ${\rm RM}_{\rm sh,tot}$ are different for the CC and merging of two WDs cases. For $n=10$ and CC channel ${\rm DM}_{\rm sh,tot} \propto t^{-0.88}$ and ${\rm RM_{\rm sh,tot}}\propto t ^{-1.88}$ during the free expansion while ${\rm DM_{\rm sh,tot}}\propto t^{-0.67}$ and ${\rm RM_{\rm sh,tot}}\propto t ^{-1.67}$ during the ST phase. In the WD merging case, ${\rm DM_{\rm sh,tot}}\propto t^{0.7}$ and ${\rm RM_{\rm sh,tot}}\propto t ^{0.4}$ during the free expansion and ${\rm DM_{\rm sh,tot}}\propto t^{0.4}$ and ${\rm RM_{\rm sh,tot}}\propto t ^{-0.2}$ in the ST phase. That is, while in the CC case both DM and RM always decrease, in the case of a merger the behaviour of DM and RM is more complex.  
We examine the case for $n=6$ as well because for this value of the power-law index the RM in the free expansion phase becomes a constant for the merger channel (see eq. \ref{eq:RM_FE_fw_rev} and figure \ref{fig:merger_DM_RM_n6}). However, the DM in this phase still increases with time and is $\propto t^{0.5}$. For the CC channel and $n=6$ ${\rm DM} \propto t^{-0.75}$ and ${\rm RM}\propto t ^{-1.75}$ (see figure \ref{fig:CC_DM_RM}) during free expansion. As the evolution of DM and RM in the ST phase does not depend on $n$ the time dependence of DM and RM in this phase are same for both the values of $n$ considered for the merger models. The same is also true for CC channel.
Assuming an uniform density profile for the ejected material \citet{piro18} obtained DM and RM both $\propto t^{-0.5}$ (DM $\propto t^{-1.5}$ and RM $\propto t^{-2.0}$) for a constant density (wind) medium in the free expansion phase. However, in the ST phase their estimates are similar to what we have obtained since the reverse shock in this phase ploughs through the flat part of the ejecta.

\section{Conclusions}
\label{sec:conclusion}
The evolution of the DM and RM in the case of the double WD mergers are unique in nature and different from the CC case. As the ejecta are  between $0.001$ and $0.3$ $\msun$, they would become transparent to the radio signals within a couple of tens of days after the merger. Moreover, due to the high temperature ($\gsim 10^{6}$ K) of the shocked shell and the low density of the CSM ($n_{\rm ISM} \lsim 100$ $\ccc$) these media are optically thin to radio pulses at 1 GHz right from the beginning. The ionised ejecta dominate the DM evolution in the first few years up to couple of decades depending on the density of the CSM. Beyond this the shocked shells influence the evolution. In the case of merger scenario with $n=10$ the DM and RM from the shocked shell initially in the free expansion phase increase with time. For $n=6$ while the DM shows a similar trend the RM does not change during this phase.
However, for both the values of $n$, as the ejecta move into the ST phase while the DM still grows with time the RM diminishes. Interestingly, this ST feature of WD merger is somewhat similar to that of the repeater FRB 121102. 
For FRB 121102 the observed DM is $\sim$ 560 pc $\ccc$ \citep{spitler14,spitler16}, which has shown an increase by 1-3  pc $\ccc$ in the last four years \citep{hessels19,josephy19}, whereas the RM has decreased from around $1.46 \times 10^{5}$ ${\rm rad~ m^{-2}}$ to $1.33 \times 10^{5}$ ${\rm rad~ m^{-2}}$ over seven months \citep{michilli18}. For WDcaseA it is found that for a CSM density of 50 $\ccc$ the shocked shell can contribute $\sim 10^{5}$ rad m$^{-2}$ to the RM. This is represented with green circles in the upper right panel of figures \ref{fig:merger_DM_RM_n10} and \ref{fig:merger_DM_RM_n6}. In case of WDcaseB this high value of RM is attainable only for $n=6$ in the free expansion phase (see right middle panel of figure \ref{fig:merger_DM_RM_n6}).
In these models the decline rate of the RM is much slower in the ST phase compare to what has been observed for FRB 121102. However, a faster decrease is expected in case the ejecta are evolving from the free expansion to ST phase. That is why we place the FRB 121102 near the transition point in figure \ref{fig:merger_DM_RM_n6}. 
It is noteworthy that the detected DM variation is consistent with the WD merger scenario in the ST phase. Therefore, it is possible that the source of repeater FRB 121102 was a product of the merger of two WDs. 
 The advantage of the merging hypothesis is that mergers can occur in all types of galaxies, regardless of their age, morphology and metallicities due to the large delay times required by gravitational radiation driven processes.

\par 
Though the repeater FRB 121102 resides in a star forming dwarf galaxy and is associated with a persistence radio source \citep{chatterjee17,eftekhari17} the localisation of three non-repeater bursts, FRB 180924, FRB 190523 and FRB\,181112 reveal that they dwell in passive galaxies, which are $\sim 100$ times massive compare to the host galaxy of FRB 121102 \citep{bannister19,ravi19,prochaska19}. Recently, 9 new repeaters have been detected by CHIME \citep{Andersen2019} and ASKAP \citep{kumar19} ( 8 of them are from CHIME  and 1 from ASKAP). 
There seems to be some evidence that the bursts observed in repeaters are wider than those from non-repeaters indicating that different emission mechanisms may be at play. The DM of the 8 CHIME repeaters found to be constant over an observing period of around 6 months. However, except one, FRB\,180916.J0158+65, with an absolute RM value of $114.6 \pm 0.6$ ${\rm rad~ m^{-2}}$, for none other bursts the RM is known. This FRB is displayed with green triangles in the right panels of Figures \ref{fig:merger_DM_RM_n10}, \ref{fig:merger_DM_RM_n6} and \ref{fig:CC_DM_RM}. In case of WD merger scenario and $n=6$ as RM is constant in the free expansion phase this FRB is placed in the midway of cyan line in the upper right plot of Figure.\ref{fig:merger_DM_RM_n6}. For other models, WDcaseB and WDcaseC, and $n=6$ a $n_{\rm ISM}$ between 1 $\ccc$ and 0.1 $\ccc$ can account for the observed RM of FRB\,180916.J0158+65. The same is true for WDcaseB when $n=10$. The constant DM of the new CHIME repeates are consistent with our merger models as the rate of change of DM in our models are slow, and, therefore, to measure an significant change one would require at least a few couple of years of time. In case of CC this rate of change of DM is faster compare to the merger scenario, however to see substantial difference one might require to wait for few years. 
For the ASKAP repeater a slight decrease in DM has been observed within a time span of two years. As a result, this FRB seems to favour the CC channel. 
With the current generation of telescopes, i.e., ASKAP \citep{mcconnell16}, CHIME \citep{CHIME18}, and other wide-field radio telescopes it is expected that the number of repeating FRBs will increase significantly in the near future. From the high volumetric rate of FRBs, \citet{ravi19b} estimate that the majority of FRBs may repeat during their lifetimes. An investigation of the time evolution of DM and RM of these bursts will then enable us to acquire more information about the progenitors of the repeating FRBs, and if some of them are due to the merger of WDs the signature of the merger will be noticeable in that study.      

\section*{Acknowledgements}
 E.K acknowledges the Australian Research Council (ARC) grant DP180100857.  

\vspace{-0.4cm}
\bibliographystyle{mnras}
\bibliography{referns} 

\bsp	
\label{lastpage}
\end{document}